\documentclass[journal]{IEEEtran}
\usepackage{ifpdf}
\usepackage{subfigure}
\usepackage{times}
\usepackage{soul}
\usepackage{url}
\usepackage[hidelinks]{hyperref}
\usepackage[utf8]{inputenc}
\usepackage[small]{caption}
\usepackage{graphicx}
\usepackage{amsmath}
\usepackage{amsthm}
\usepackage{booktabs}
\usepackage{algorithm}
\usepackage{algorithmic}
\usepackage{mathtools}   
\usepackage[T1]{fontenc}
\usepackage{amssymb}
\usepackage{stackengine}
\usepackage{empheq}
\usepackage{enumitem}
\usepackage{bm}
\usepackage{makecell}
\usepackage{multirow}

\usepackage{cite}

\ifCLASSINFOpdf
\else
\fi
\usepackage{amsmath}
\usepackage{algorithmic}

\usepackage{array}

\usepackage{stfloats}
\newcommand{\etal}{\textit{et al.}}

\begin{document}
	%
	\title{Can Machines Generate Personalized Music? A Hybrid Favorite-aware Method for User Preference Music Transfer}
	%
	%
	%
	
	\author{Zhejing~Hu,~Yan~Liu,~Gong~Chen,~and~Yongxu~Liu
		}

	\maketitle
	
	
	\begin{abstract}
    User preference music transfer (UPMT) is a new problem in music style transfer that can be applied to many scenarios but remains understudied. Transferring an arbitrary song to fit a user’s preferences increases musical diversity and improves user engagement, which can greatly benefit individuals’ mental health. Most music style transfer approaches rely on data-driven methods. In general, however, constructing a large training dataset is challenging because users can rarely provide enough of their favorite songs. To address this problem, this paper proposes a novel hybrid method called User Preference Transformer (UP-Transformer) which uses prior knowledge of only one piece of a user’s favorite music. Based on the distribution of music events in the provided music, we propose a new favorite-aware loss function to fine-tune the Transformer-based model. Two steps are proposed in the transfer phase to achieve UPMT based on the extracted music pattern in a user’s favorite music. Additionally, to alleviate the problem of evaluating melodic similarity in music style transfer, we propose a new concept called pattern similarity (PS) to measure the similarity between two pieces of music. Statistical tests indicate that the results of PS are consistent with the similarity score in a qualitative experiment. Furthermore, experimental results on subjects show that the transferred music achieves better performance in musicality, similarity, and user preferences.
    		
	\end{abstract}
	\begin{IEEEkeywords}
		Music style transfer; Artificial music intelligence; Automatic music generation
	\end{IEEEkeywords}
	
	\section{Introduction}
	
\begin{figure}
		{\includegraphics[width=\linewidth,trim=0 0 0 20,clip]{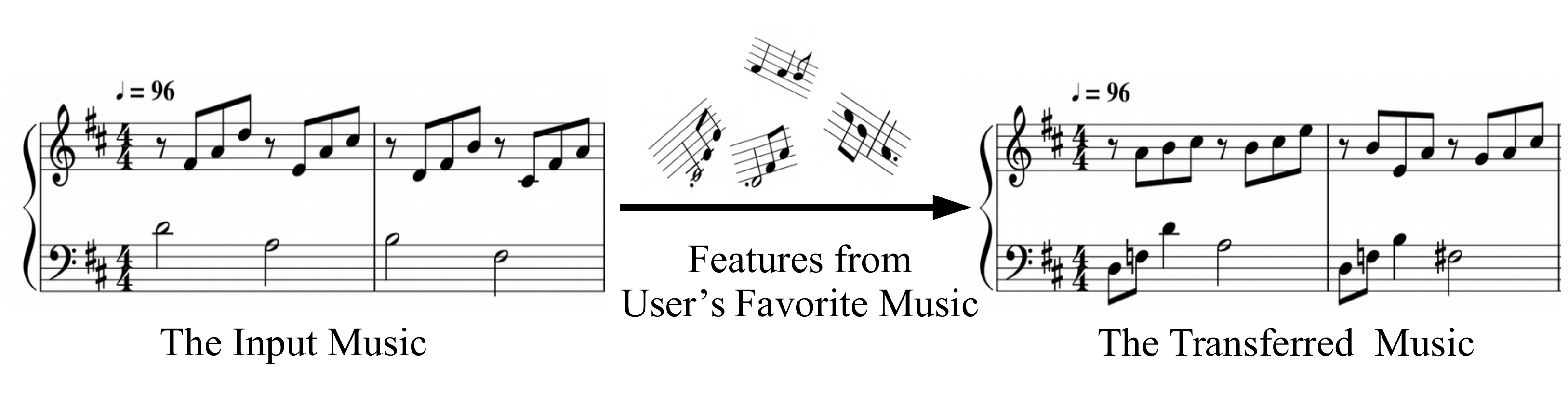}}
		\caption{A demonstration of UPMT: Transferring symbolic input music to new symbolic music that fits a user’s preferences based on features of their favorite music.}
		\label{fig11}
		
	\end{figure}

	\IEEEPARstart{T}{here} has been recent growth in research around music style transfer, a technique that transfers the style of one piece of music to another based on different levels of music representations \cite{dai2018music}. Music style transfer is considered important because it increases music variety by reproducing existing music in a creative way. Music style transfer also helps humans compose music more easily and makes music composition more accessible to all \cite{10.5555/3367471.3367696}.
	To date, only a few works have addressed music transfer based on user preferences \cite{hu2020make}. We name this problem user preference music transfer (UPMT). UPMT aims to transfer symbolic input music to a new piece of symbolic music that meets a user’s preferences based on features of their favorite music (Fig.\ref{fig11}). UPMT can be examined in numerous applications, such as anxiety reduction \cite{mornhinweg1992effects,bradt2013music}, personality studies \cite{rawlings1997music,dunn2012toward}, and music recommender systems \cite{bogdanov2010content,celma2010music}. For example, in the music therapy domain, individuals have been found to be more engaged when listening to their favorite music \cite{kanehira2018enhanced}. Yet existing therapeutic music is limited and not personalized, which restricts the impact of music therapy. Transferring therapeutic music to suit user preferences can lead to more effective anxiety treatments that alleviate symptoms. This technique also avoids copyright issues. 
	
	Currently, music transfer tends to involve data-driven methods such as VAE-based \cite{brunner2018midi,yang2019inspecting} and GAN-based models \cite{brunner2018symbolic,lu2019play}, which require a large number of data samples for model training. However, one user can typically only provide a limited number of their favorite songs, which is insufficient for training. A small number of data samples further evokes the challenge of representing the features of users' favorite music, resulting in unpleasant output and inferior performance. 
	
	To alleviate the problem of limited training data, studies in fields such as image classification and recommender systems have proposed hybrid methods that introduce knowledge into data-driven methods to improve model explainability and performance \cite{10.5555/3172077.3172109,garcez2019neural,Marino_2017_CVPR,donadello2018semantic}. For example, Marino \etal \cite{Marino_2017_CVPR} used prior semantic knowledge in the form of knowledge graphs to improve image classification performance. Donadello \etal \cite{donadello2018semantic} extracted semantic representations in a knowledge base to enhance the quality of recommender systems. Despite these advances, the approaches cannot be directly applied to music, as the knowledge structure differs in this domain. With respect to music specifically, Liu \etal \cite{liu2020incorporating} proposed incorporating music knowledge such as $\text{18}^{\text{th}}$-century counterpoint rules to augment data for music generation; however, they focused on producing more data for training rather than improving the computational model itself. In addition, a knowledge base is usually needed to facilitate training. Constructing a knowledge base is difficult in UPMT because only a limited number of music samples can be used.
	
	Inspired by existing hybrid approaches, we propose a hybrid music transfer method named User Preference Transformer (UP-Transformer). UP-Transformer utilizes the music knowledge of only one piece of a user’s favorite music and adopts the pre-trained Transformer-XL \cite{dai-etal-2019-transformer} as the backbone model to address the UPMT problem in two phases: a fine-tuning phase and a transfer phase. In this paper, music knowledge refers to prior knowledge of a user’s favorite music that represents the distribution of music events and a unique music pattern called the signature music pattern interval (SMPI). A favorite-aware loss function is proposed to fine-tune the pre-trained Transformer-XL \cite{dai-etal-2019-transformer} in the fine-tuning phase. In the transfer phase, an event selection step and an event learning step are executed to ensure that the output music is enjoyable and fits the user’s preferences. In an experiment, we design a new metric called pattern similarity (PS) to measure music PS, which enriches quantitative experiments involving music style transfer. A qualitative experiment indicates that the output music achieves better performance in terms of musicality, similarity, and preference. These results suggest that the proposed method can successfully transfer input music to fit a user’s preferences based on only one piece of favorite music.

	The novelty and contributions of this paper are as follows:
	\begin{enumerate}
		\item We propose a novel hybrid method named UP-Transformer by using music knowledge based on a user’s favorite music and Transformer-XL. UP-Transformer provides a new scheme of transferring music by altering and fixing certain music events in the input symbolic music, which can then be applied to different transfer tasks.
				
		\item We define a new favorite-aware loss function to fine-tune the backbone Transformer-XL along with two novel steps to transfer music. The proposed method mitigates the problem wherein data-driven methods cannot be well trained on small datasets. This method also significantly improves output quality by better suiting user preferences. 
		
		\item We propose a new metric to measure the melodic similarity between two music sequences. We then conduct a statistical experiment on randomly selected music samples. The experiment demonstrates the effectiveness and feasibility of the new metric, which can be further applied to different music tasks to measure the melodies in music.
	\end{enumerate}
	
	The rest of this paper is organized as follows. In Section II, we review related work. We detail the proposed model in Section III. Section IV presents our implementation process and experimental results. Conclusions and directions for future work are summarized in Section V.

	\section{Related Work}
	In this section, we review existing work related to our proposed method, which can be divided into two threads: music style transfer and Transformer-based methods of music generation.
	\subsection{Music Style Transfer}
	The term ``style transfer'' was proposed by \cite{gatys2015neural}, who transferred the style of an input image to another specific style. A large body of work has since emerged in this field \cite{jing2019neural}. Music style transfer applies the principle of image style transfer in which input is transferred to a new style while retaining content-related information. However, different from image style transfer, the representation methods, transfer methods, and transfer types are more varied and complicated in music style transfer.  
	
	In terms of representation methods, some researchers have transferred a music style directly onto a raw audio signal \cite{lu2019play,ye2020music,huang2018timbretron}, which is a sound approach to representation and can be depicted as waveforms and spectrograms. Others have aimed to transfer a music style based on MIDI-format piano-roll \cite{brunner2018midi,yang2019inspecting,yang2019deep,malik2017neural}, a performance control representation method which features symbolic scores. MIDI-format piano-roll can be further depicted as a MIDI-like event matrix and beat-based event sequence. In this paper, audio music is not the scope of our research and we use beat-based event sequence to represent symbolic music since it is denser and contains more music information, such as tempo changes \cite{huang2020pop}, which can be studied in greater depth in music style transfer. 
		
	With respect to transfer methods, some scholars have proposed disentangling the content and style features of music \cite{yang2019deep,yang2019inspecting,lu2019play,cifka2020groove2groove,hu2020make}. By minimizing content and style feature differences, the content features of new music should be similar to the content music while the style features of new music should be similar to the style music. For instance, Hu \etal \cite{hu2020make} attempted to transfer music to be therapeutic based on one piece of a user’s favorite music. Yet ``music style'' is a notoriously fuzzy term, making it difficult to completely strip style information from input music. Others have tried to directly transfer music from one style to another without differentiating between content and style features \cite{lu2018transferring,brunner2018symbolic,pasini2019melgan,ondrej_cifka_2019_3527878}. Overall, to avoid disentangling content and style features and improve interpretability, the proposed method provides a new way to transfer the music by altering and fixing some music events of the input symbolic music.
	
	Regarding style transfer, most music style transfer methods focus on transferring music genres \cite{brunner2018symbolic,lu2018transferring}, music timbres \cite{huang2018timbretron,musictranslation}, and composition styles \cite{10.5555/3367471.3367696}. Although these approaches have achieved satisfactory performance, they are generally trained on large music samples and do not demonstrate the favorite-aware capacity to suit a user’s personal tastes. It would therefore be inappropriate to apply these models to UPMT. Recently, C\'{i}fka \etal \cite{cifka2020groove2groove} proposed one-shot music style transfer that only requires one piece of music but entails a complicated process to generate parallel training examples for supervised learning; this procedure is time consuming, and the quality of synthetic data is uncertain. Therefore, to better fit the user's favorite music, we propose the favorite-aware mechanism including one favorite-aware loss function and two knowledge based steps to learn unique user preferences.
	
	\subsection{Transformer-based Methods} 
	The Transformer \cite{vaswani2017attention} has achieved satisfactory results in different generation tasks that require maintaining long-range coherence. The unique beat-bar hierarchical structure and long-range repetition of music suggest that the Transformer-based model might lead to better performance in music modeling. Transformer-based methods have recently been applied to music problems \cite{huang2018music,choi2020encoding,jiang2020transformer,park2019bi}. Huang \etal \cite{huang2018music} proposed Music Transformer to generate music and were the first to apply Transformer to music generation. They showed that a Transformer-based model can generate qualified music. However, they used MIDI-like representation methods to represent music. This method misses musical features such as tempo changes and fails to measure note duration explicitly, which further influences the performance in terms of rhythm, an important aspect of music. Different Transformer-based approaches \cite{choi2020encoding,jiang2020transformer,park2019bi} were subsequently proposed and have achieved better performance. Two representative methods are the Pop music Transformer (PopMT) \cite{huang2020pop} and PopMAG \cite{ren2020popmag}. In PopMT, the authors proposed a new representation method called REMI and applied Transformer-XL \cite{dai-etal-2019-transformer} as the backbone to generate pop music to improve output performance. In PopMAG, the authors presented another novel music representation method called MuMIDI and addressed the challenge of multi-track harmony modeling. To the best of our knowledge, these methods are data-driven and are not designed for music transfer. The proposed approach applies Transformer-based methods to accomplish music style transfer by fixing and modifying certain music events at each time step.
	
	\section{Proposed Approach}
	In this section, the proposed UP-Transformer will be introduced in detail. We first formulate the UPMT task. Then we present an overview of our proposed approach and describe the fine-tuning phase and transfer phase.
	
    \begin{figure*}[t]
		\centering
		\includegraphics[width=\linewidth,trim=10 0 60 0,clip]{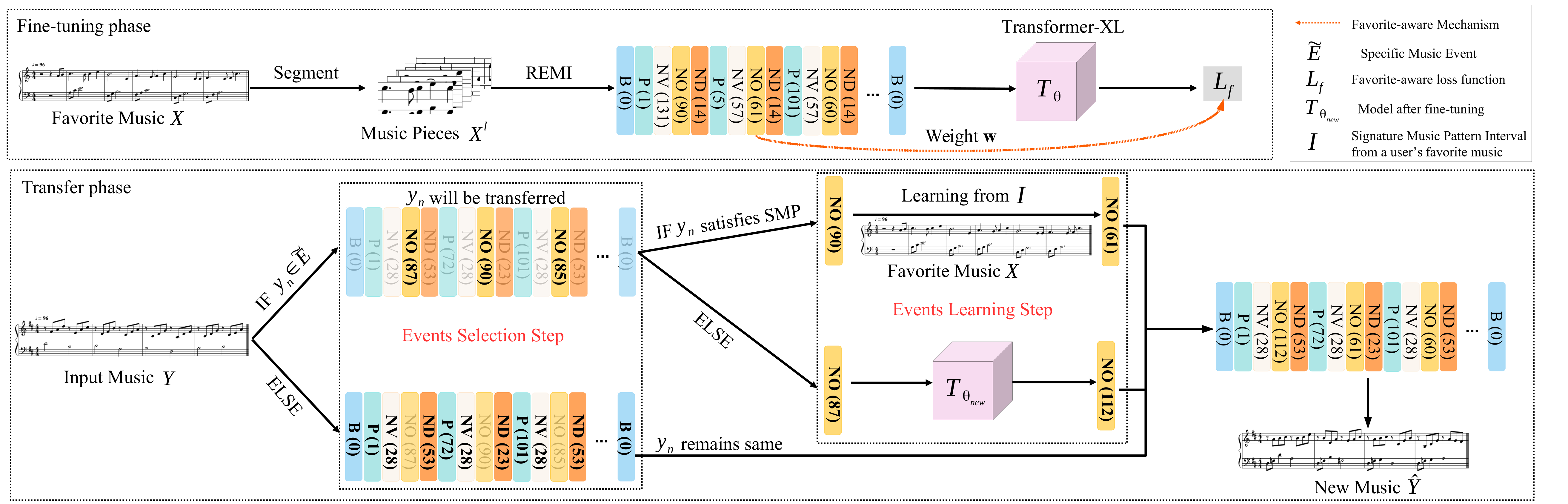}
		\caption{A general framework of the proposed model. The model consists of two phases, a fine-tuning phase and a transfer phase. SMP: signature music pattern. In this demonstration, only \textit{Note On} events are transferred. $\hat{Y}$ contains information such as the number of bars, number of notes, note duration, and note position from $Y$ based on the event selection step but learns note-on information based on the event learning step.}
		\label{fig1}
	\end{figure*}

	\subsection{Problem Formulation}
	Considering one piece of a user’s favorite music $X$ that is provided by the user, prior knowledge (e.g., the distribution of different music events and unique music patterns) can be extracted from a user’s favorite music $X$, which is denoted as $K(X)$. The goal of UPMT is that, given an arbitrary piece of music $Y$, a method $M$ with parameters $\phi$ can be used to transfer music so that the transferred new music $\hat{Y}$ contains knowledge $K$ of a user’s favorite music $X$ and preserves important information (e.g., music structure and length) from the original music $Y$. This process can be formulated as follows:
	\begin{equation}
		\hat{Y} = M(Y,K(X);\phi)
	\end{equation}
	
	\subsection{Model Overview}
	To address the aforementioned data limitation problem and enhance model performance, we propose UP-Transformer. This model utilizes music knowledge, including the distribution of music events and SMPI in a user’s favorite music, and adopts a pre-trained Transformer-XL \cite{dai-etal-2019-transformer} as the backbone model to transfer music. A diagram of the proposed framework appears in Fig \ref{fig1}. The framework includes two phases: a fine-tuning phase and a transfer phase.
	
	Each phase proceeds as follows:
	\begin{enumerate}
		\item The user’s favorite music $X$ is segmented into many pieces. 
		\item An up-to-date music representation method called REMI \cite{huang2020pop} is applied to transfer symbolic music data into a music event sequence, which will be discussed in Section III.C. 
		\item A pre-trained Transformer-XL model is used as the backbone model and will be fine-tuned based on a novel favorite-aware loss function that utilizes distribution knowledge from the user’s favorite music, to be addressed in Section III.D. 
		\item In the transfer phase, two steps (i.e., event selection and event learning) will be adopted along with SMPI from the user’s favorite music and the pre-trained Transformer-XL to transfer input music into a new piece of music and satisfy user preferences; this task will be described in Section III.E.
	\end{enumerate}
	
	The novelties of UP-Transformer are twofold. First, we propose a new favorite-aware loss function to fine-tune Transformer-XL based on the distribution of different music events in one piece of a user’s favorite music. Second, two music steps are implemented to regulate the transfer process based on the extracted SMPI.

	\subsection{Event-based Music Representation}
	
\begin{figure}[t]
	\centering
	\includegraphics[width=\linewidth,trim=30 260 40 110,clip]{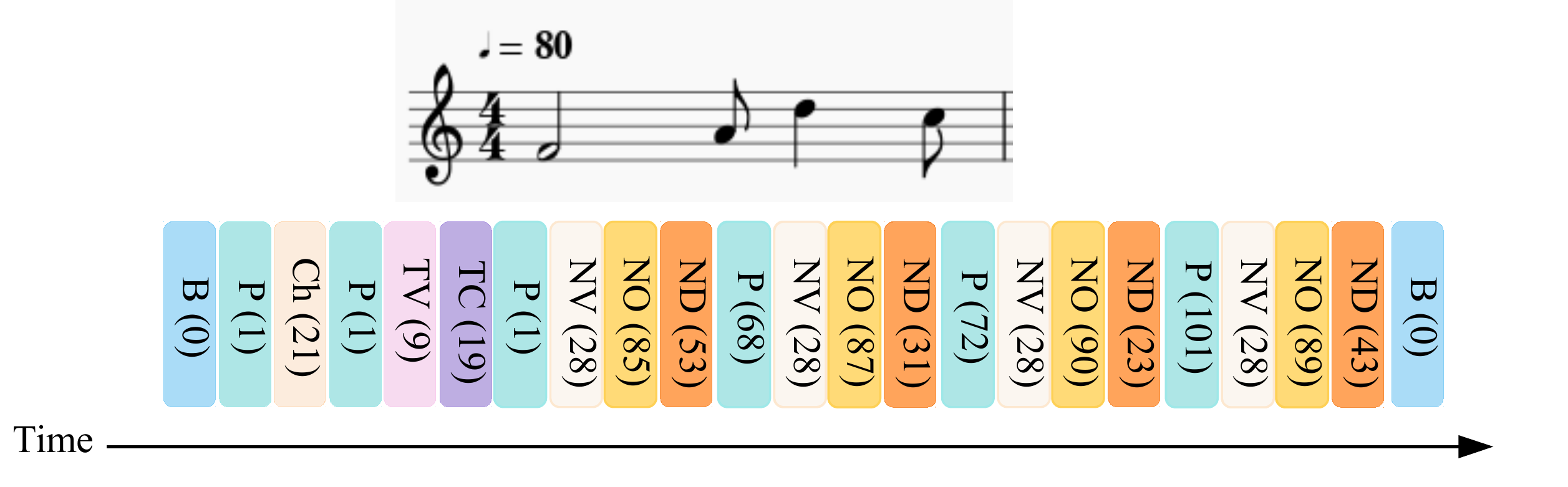}
	\caption{An example of encoding symbolic music into a sequence using REMI.}
	\label{remi}
\end{figure}

	In this paper, REMI \cite{huang2020pop} is used to represent MIDI scores in symbolic music. In REMI, there are eight different music events, which are defined based on note, tempo and chord information of symbolic music. Each music event includes one or more classes to reflect different values. For example, \textit{Bar (B)} indicates the bar line in piano rolls. One note contains four different events, \textit{Position (P)}, \textit{Note Velocity (NV)}, \textit{Note On (NO)} and \textit{Note Duration(ND)}, respectively. \textit{P} represents one of the $q$ possible discrete locations in a bar, where $q$ indicates the time resolution adopted to represent a bar. \textit{NV} is quantized into 32 levels to indicate the perceptual loudness of the note event. A collection of 128 \textit{NO} events indicates the onset of MIDI pitches from 0 to 127 ($C_{-1}$ to $G_{9}$); that is, there are 128 classes in the \textit{Note On} event.  \textit{ND} is used to indicate the length of one note, which contains a value ranging from 1 to 64. In addition, \textit{Tempo} event indicates local tempo changes and is added at every beat. \textit{Tempo Class (TC)} and \textit{Tempo Value (TV)} are used to represent local tempo values between 30 and 209 beats per minute (BPM). \textit{Chord (Ch)} events are also considered in REMI (i.e., 12 chord roots and five chord qualities). Each \textit{Tempo Class}, \textit{Tempo Velocity}, and \textit{Chord} is preceded by a \textit{Position} event. Figure \ref{remi} demonstrates the music tokenization process.

	Compared to the MIDI-like event matrix, REMI \cite{huang2020pop} explicitly defines different events in sequence. REMI thus includes more information, such as tempo changes and chord information, and produces a denser representation. Further, each note is grouped into four consecutive events – \textit{P}, \textit{NV}, \textit{NO}, and \textit{ND} – which affords us more flexibility in manipulating the sequence. 
	
	\subsection{UP-Transformer in Fine-tuning Phase}
	
	\begin{algorithm}
		\caption{UP-Transformer in fine-tuning phase.}
		\label{algorithm1}
		\textbf{Input}: User’s favorite music $X$, a pre-trained Transformer-XL network $T_{\theta}$\\
		\textbf{Output}: Optimized Transformer-XL network $T_{\theta_{new}}$
		
		\begin{algorithmic}[1] 
			\WHILE{ not converge}
			\STATE Compute weight $w_{c}$ with Equation \ref{eqs1};
			\STATE Train $T_{\theta}$ using Equation \ref{eqs2}.
			\ENDWHILE
			\STATE \textbf{return} $T_{\theta_{new}}$
		\end{algorithmic}
	\end{algorithm}
	
	Previous work \cite{donahue2019lakhnes} has shown that Transformer-based models can benefit from pre-training techniques on large-scale music datasets. Therefore, rather than using CNN-based style transfer techniques from the image domain \cite{park2019arbitrary,hsieh2019one}, we begin with a Transformer-XL \cite{dai-etal-2019-transformer} model with a default attention mechanism that was pre-trained on a pop music dataset. Transformer-XL introduces the notion of recurrence and revises the positional encoding scheme which, in theory, can encode arbitrary long content into a fixed length. In addition, to fully leverage a user’s favorite music and to guide the model to learn specific music events in the case of limited training samples, we propose building a novel favorite-aware loss function that directly learns the music event distribution from a user’s favorite music \cite{huang2020pop}.
	
	After REMI \cite{huang2020pop}, each token is a class (value), and each class belongs to one of the eight music events described in the previous sub-section. Each token therefore correlates with one of these eight music events. In the original Transformer-XL \cite{dai-etal-2019-transformer}, the model is optimized using the cross-entropy loss function, which treats all classes equally and assigns each class the same weight. However, the physical meaning of each music event varies. For instance, notes constitute the most important information in music. In a sequence, the combination of \textit{P}, \textit{NV}, \textit{NO}, and \textit{ND} represents one note. It is thus reasonable to pay closer attention to note-related events than to other events such as \textit{TC} and \textit{TV}. 
	
	Inspired by the physical meaning of each music event and due to limited training samples, the favorite-aware loss function gives greater attention to music events that have more effects on a user’s favorite music. In other words, favorite awareness is calculated based on the distribution of events in the user’s favorite music. The fine-tuning procedure is detailed in Algorithm \ref{algorithm1}. We also conduct experiments to validate this model design.
	
	Formally, let $E=\{e_{B},e_{P},e_{Ch},e_{TC},e_{TV},e_{NV},e_{NO},e_{ND}\}$ be the set of eight music events. We use $e_{k}$ and $c$ to represent an arbitrary music event and its corresponding class, respectively. Here, $e_{k} \in E$ is also a set that contains one or more classes and $c \in e_{k}$.
	
	Given one music piece $X^{l}$ with sequence length $N$, $X^{l}=\{x_{1}^{l},x_{2}^{l},\ldots,x_{N}^{l}\}$ where $x_{n}^{l}=c \in e_{k}$, then Transformer-XL \cite{dai-etal-2019-transformer} $T_{\theta}(\cdot)$ can predict the next token $\hat{x}^{l}_{i}$ conditioned on the previously generated tokens $x_{t<i}^{l}$, formulated as
	$\hat{x}^{l}_{i} = T_{\theta}(x_{t<i}^{l})$. 
	The favorite-aware loss $L_{f}$ is defined as
	\begin{equation}
		L_{f}=  -\frac{1}{N} \sum_{i=1}^{N} \sum_{c=1}^{C} w_{c} * p_{i,c} log(q_{i,c}) ,
		\label{eqs2}
	\end{equation}
	where $N$ is the sequence length, $C$ is the total number of classes, $p_{i,c}$ corresponds to the $c^{\text{th}}$ element of the one-hot encoded label of the ground truth $x_{i}^{l}$, and $q_{i,c}$ is the predicted category probability that observation $i$ is of class $c$. $w_{c}$ is the $c^{\text{th}}$ favorite-aware weight $\textbf{w}=\{w_{1},w_{2},\ldots,w_{C}\}$ and can be further calculated from a user’s favorite music $X$ with the following formula:
	
	\begin{equation}
		w_{c} =\left\{
		\begin{array}{ll}
			\alpha+\dfrac{Count(c)}{\sum_{c \in \tilde{E}} Count(c)},\qquad\ \ \ \ \ \ if \ c \in \tilde{E},\\
			\alpha,\qquad\qquad\qquad\qquad\qquad\qquad\qquad o.w.
		\end{array}
		\right.
		\label{eqs1}
	\end{equation}
	where $\tilde{E} \in E$ denotes selected music events, $Count(\cdot)$ is a function that counts the appearance of an input element in the sequence $X$, and $\alpha>0$ is a modifiable parameter. In this way, information that appears more frequently in the user’s favorite music $X$ will have a higher weight. For events that are not selected for transfer, the same weight will be given with a fixed value $\alpha$.
	
	\subsection{UP-Transformer in Transfer Phase}
	
	\begin{algorithm}[tb]
		\caption{UP-Transformer in transfer phase.}
		\label{alg:algorithm2}
		\textbf{Input}: The user’s favorite music $X =\{x_{1},x_{2},\ldots,x_{N_{x}}\}$, an arbitrary piece of music $Y =\{y_{1},y_{2},\ldots,y_{N_{y}}\}$ \\
		\textbf{Parameter}: $\theta_{new}$\\
		\textbf{Output}: The updated sequence $\hat{Y} = \{\hat{y}_{1},\hat{y}_{2},\ldots,\hat{y}_{N_{y}}\}$
		
		\begin{algorithmic}[1] 
			\STATE Find all $\tilde{E}$ in $X$ and denote as $\tilde{E}^{X}= \{\tilde{e}_{1}^{X},\tilde{e}_{2}^{X},\ldots,\tilde{e}_{z_{x}}^{X}\}$
			\STATE Extract an SMP $M = \{m_{1},m_{2},\ldots, m_{a}\}$ from $\tilde{E}^{X}$ and denote it as music pattern interval $I = \{i_{1},i_{2},\ldots,i_{a-1}\}$ 
			\STATE Let $v=0$, flag$=0$,  $n=1$ and $idx=1$
			\WHILE{$n<N_{y}$}
			\IF {$y_{n} \in \tilde{E}$}
			\IF {flag$=0$}
			\STATE $\hat{y}_{n} = T_{\theta_{new}}(\hat{y}_{1},\ldots,\hat{y}_{n-1})$
			\IF {$\hat{y}_{n} - \hat{y}_{v} = i_{1}$}
			\STATE flag=1
			\ENDIF
			\ELSE
			\STATE $\hat{y}_{n}$ = $\hat{y}_{v}+i_{idx}$
			\STATE $idx =idx+1$
			\IF {$idx>a-1$}
			\STATE $idx=1$
			\STATE flag$=0$
			\ENDIF
			\ENDIF
			\STATE $v=n$
			\ELSE
			\STATE$\hat{y}_{n}=y_{n}$
			\ENDIF
			\STATE $n=n+1$
			\ENDWHILE
			\STATE \textbf{return} $\hat{Y}$
		\end{algorithmic}
	\end{algorithm}
	
	\begin{algorithm}[tb]
		\caption{SMP extractor}
		\label{alg:algorithm_smp}
		\textbf{Input}: $\tilde{E}^{X}$\\
		\textbf{Parameter}: $a$\\
		\textbf{Output}: SMP $M$
		\begin{algorithmic}[1] 
			\STATE Generate a dictionary $dict$ for all sub-lists of length $a$ in $\tilde{E}^{X}$.
			\IF {Repeated patterns exist}
			\STATE Find all repeated patterns in $dict$ and assign them  to an empty list $R$.
			\STATE $M$ = Randomly select one repeated pattern from $R$.
			\ELSE  
			\STATE No repeated patterns, please shorten the length $a$.
			\ENDIF
			\STATE \textbf{return} $M$ 
		\end{algorithmic}
	\end{algorithm}
	
	In the transfer phase, an arbitrary input sample $Y = \{y_{1},y_{2},\ldots,y_{N_{y}}\}$ with length $N_{y}$ will be transferred to a new piece of music $\hat{Y}=\{\hat{y}_{1},\hat{y}_{2},\ldots,\hat{y}_{N_{y}}\}$ based on the user’s favorite music $X = \{x_{1},x_{2},\ldots,x_{N_{x}}\}$ with length $N_{x}$. We apply two music steps to increase the model’s explainability and improve its performance. 
	
	Specifically, the event selection and event learning steps are included to solve UPMT. The details of these two steps appear in Algorithm \ref{alg:algorithm2}. First, we define the event selection step as follows: only selected music events $\tilde{E} \in E$ in $Y$ are transferred while other music events remain the same. The event selection step can be executed as follows:
	\begin{enumerate}
		\item If $y_{n} \in \tilde{E}$, then $y_{n}$ will be transferred and $\hat{y}_{n}$ will be predicted by implementing the event learning step (lines 5–19).
		\item If $y_{n} \notin \tilde{E}$, then  $\hat{y}_{n} = y_{n}$ (lines 20-22).
	\end{enumerate}

	For example, if we define $\tilde{E} = e_{NO}$, then only the \textit{NO} event in $Y$ will be transferred. Other events will remain the same. Event selection fixes music events that we do not want to transfer and only focuses on a specific music event to ensure that the new music follows the structure of the input music, thus preserving information from the original music $Y$. 
	
	The event learning step controls the predicted event by forcing it to be similar to the user’s favorite music under certain conditions. Melody is one of the most basic elements in music. Melodic sequences, melodic phrases, or motifs can be considered music patterns and are represented as groups of consecutive notes that make sense together. In music, at least one music pattern always appears repeatedly that makes the melody unique. We define this music pattern as SMP because it is akin to a signature that makes music recognizable. To identify this pattern, we first find all selected music events $\tilde{E}$ in a user’s favorite music $X$ and denote these events as $\tilde{E}^{X}=\{\tilde{e}_{1}^{X},\tilde{e}_{2}^{X},\ldots,\tilde{e}_{z_{X}}^{X}\}$, where $z_{X}$ is the total number of selected music events in $X$. We then extract SMP from $\tilde{E}^{X}$ based on a method called SMP Extractor (Algorithm \ref{alg:algorithm_smp}). We define the SMP as $M = \{m_{1},m_{2},\ldots,m_{a}\}$ and $M \subset \tilde{E}^{X}$, where $a$ is the length of SMP that can be changed based on different users’ favorite music.  
	
	In addition, rather than applying SMP directly, we calculate the difference and call it SMPI $I=\{i_{1},i_{2},\ldots,i_{a-1}\}$, which is calculated as $I = \{m_{2}-m_{1},m_{3}-m_{2},\ldots,m_{a}-m_{a-1}\}$. For example, if the selected event $\tilde{E}=e_{NO}$ and an SMP $M$ is $\{60,62,62,64,62,62,60,68\}$, then SMPI $I$ is represented as $\{2,0,2,-2,0,-2,8\}$. We use SMPI $I$ instead of SMP $M$ because music tends to have different pitch levels, but in terms of music similarity, relative differences are more representative; humans barely hear the difference if the ratio of two notes’ frequencies (i.e., SMPI) is the same \cite{meyer1996style}. For instance, an SMP $M=\{72,74,74,76,74,74,72,80\}$ sounds familiar to users compared with SMP $M=\{60,62,62,64,62,62,60,68\}$ where $I$ remains the same.
	
	The event learning step is described in Algorithm 2 (lines 6–19) and can be executed as follows:
	\begin{enumerate}
		\item A $flag$ is set to indicate if the event learning step will be executed or not. 
		\item If $flag =0$, then the event learning step will not be executed. The current token $\hat{y}_{n}$ will be predicted by the pre-trained Transformer-XL \cite{dai-etal-2019-transformer} $T_{\theta_{new}}(\cdot)$; that is, $\hat{y}_{n} =T_{\theta_{new}}(\hat{y}_{1},\ldots,\hat{y}_{n-1})$ (lines 6–7). 
		\subitem If the difference between the current predicted token and the previous predicted token $\hat{y}_{n} - \hat{y}_{v} = i_{1}$ (the difference matches the first item in SMPI $I$), then $flag$ will be set to 1 to indicate that the event learning step will be executed in the following steps (lines 8–10).
		
		\item If $flag =1$, then the event learning step will be executed. Specifically, subsequent tokens that belong to the selected event $\tilde{E}$ will not be predicted by Transformer-XL \cite{dai-etal-2019-transformer} $T_{\theta_{new}}(\cdot)$; instead, the event will be updated directly based on SMPI $I$ until all information from $I$ has been transferred to the new music $\hat{Y}$ (lines 11–13). 
		\subitem If all items in $I$ are updated to subsequent tokens, then the flag will return to 0 to indicate that updating is complete (lines 14–17).
		\item Update previous predicted token (line 19). 
	\end{enumerate}
	The event learning step first extracts SMPI from the user’s favorite music based on prior human knowledge and then regulates the result to be similar to the user’s favorite music, which improves model performance directly.

	\section{Experiments}
	In this section, we first describe our experimental settings including the dataset, implementation details, and metrics, after which our experimental results are discussed at length. 
	\subsection{Datasets} 
	All users' favorite music samples were provided by the subjects and downloaded from open sources; more specifically, each subject provided one piece of their favorite music. Twenty-three subjects were recruited. Nineteen pieces of music were included due to overlap between subjects' favorite music. In the fine-tuning phase, these 19 pieces of music were segmented into the same length. Some basic statistics of the training samples are listed in Table \ref{tab1}. 
	
	In the transfer phase, seven songs from seven genres (classical, jazz, pop, rock, folk, new age and light music) were used as input music for transfer. Detailed information about users' favorite music and the input music can be found in the supplementary materials.

	\begin{table}
		\centering
		\caption{Statistics of training data.}
		\label{tab1}
		\begin{tabular}{ccc}
			\toprule
			Statistics & Mean&  Std\\
			\midrule
			$\#$ of Segments&49.42&41.95\\
			$\#$ of \textit{Note On (NO)} &1610.21&1234.69\\						$\#$ of \textit{Note Velocity (NV)} &1610.21&1234.69\\						$\#$ of \textit{Note Duration (ND)}&1610.21&1234.69\\						$\#$ of \textit{Position (P)}&1812.73&1384.09\\						$\#$ of \textit{Bar (B)}&81.16&32.63\\						$\#$ of \textit{Tempo Class (TC)}&93.10&158.35\\						$\#$ of \textit{Tempo Value (TV)}&1.00&0.00\\						$\#$ of \textit{Chord (Ch)} &109.42&47.41\\				
			\bottomrule
		\end{tabular}
	\end{table}
	
	\subsection{Implementation Details}
	In the fine-tuning phase, because the training data were limited, one piece of a user's favorite music was divided into segments with a length of 128 tokens. In Transformer-XL, we followed the original setting mentioned in \cite{huang2020pop}. For the favorite-aware loss function, we set $\alpha = 0.01$. We fine-tuned the model for 200 epochs or until the total loss was smaller than 0.1. In the transfer phase, we chose the track containing the most note information to complete the transfer task; other tracks remained identical to the input music in order to make the music sound more similar to the original input music. After transfer, all tracks’ information was merged to generate new music.
	
	\subsection{Comparisons} 
	No existing work is directly applicable to UPMT. To evaluate the proposed method, we designed four models by extending the following related work for comparison:
	
	$\bullet$ \textbf{PopMT}\cite{huang2020pop}: This approach applies REMI to represent symbolic music data and uses Transformer-XL to generate music.
	
	$\bullet$ \textbf{PopMAG} \cite{ren2020popmag}: This approach applies MuMIDI to represent symbolic music data and uses the Recurrent Transformer encoder and decoder to generate music.
	
	$\bullet$ \textbf{CycleGAN}\cite{brunner2018symbolic}: This approach extends CycleGAN by introducing additional discriminators to transfer different music genres, which deals with matrix data.

	$\bullet$ \textbf{Therapeutic Transfer Model (TTM)} \cite{hu2020make}: This approach takes matrix data as input and transfers two pieces of music by extracting the main and secondary features using a convoluted neural network–based model.
	
	In the training phase, we first pre-trained these models based on the original training settings of each. Then, they were fine-tuned on a user’s favorite music based on their pre-training settings (e.g., sequence length, sampling parameters, and so on).
	
	In the transfer phase, PopMT and PopMAG use Transformer-XL as the backbone and were employed to complete transfer tasks by transferring specific music events $\tilde{E}$ and fixing the remaining music events, which we named PopMT\_T and PopMAG\_T. CycleGAN after fine-tuning was used to directly transfer the input music to fit users’ preferences. In TTM, we implemented the pre-trained feature extraction network used in \cite{hu2020make} to extract features. Then, the output music was optimized by assuming that the user‘s favorite music was the ``style‘‘ and the input music was the ``content’’. In addition, because the source code and parameters of PopMAG are not yet public, we tried to reproduce the model as accurately as possible.

	\subsection{Evaluation Metrics}
	Although many evaluation metrics have been proposed in various papers, quantitative evaluation in the music field remains an open question, especially in music style transfer. Following \cite{ren2020popmag}, we use the following four metrics to evaluate the transferred result and to compare our method with other models.
	
	$\bullet$ Pitch Class (\textit{P}): The distribution of pitch classes is computed.
	
	$\bullet$ Note (\textit{N}): The distribution of notes is computed.
	
	$\bullet$ Duration (\textit{D}): Duration values are quantized into 32 classes corresponding to 32 duration attributes, and the distribution of classes is computed. 
	
	$\bullet$ Inter-Onset Interval (\textit{IOI}): This element measures the interval between two note onsets. The \textit{IOI} values are quantized into 32 classes, and the distribution of classes is computed.
	
	Then, the average overlapped area (\textit{OA}) of the distributions of four factors is calculated to measure the similarity between two pieces of music:
	$$D_{A} = \dfrac{1}{N_{tracks}* N_{bars}}\sum_{i=1}^{N_{tracks}} \sum_{j=1}^{N_{bars}}OA(P^{A}_{i,j} , \hat{P}^{A}_{i,j}),$$
	where $OA$ represents the average \textit{OA} of two distributions, $P^{A}_{i,j}$ denotes the distribution of feature $A$ in the $i$-th bar and the $j$-th track in a ground truth musical piece, and $\hat{P}^{A}_{i,j}$ denotes that in the generated musical piece; $A$ can be any of four factors. It should be noted that we did not measure velocity and chord, which were evaluated in \cite{ren2020popmag}, as our focus is a transfer task in which velocity and chord information remained the same.
	
	In addition, melody is one of the most basic and distinctive features used to measure music similarity \cite{volk2012melodic}. However, the aforementioned metrics only assess music event information that does not include melodic information. To the best of our knowledge, existing measures in music style transfer do not measure melodic similarity. Therefore, we designed a new metric, PS, to assess the consecutive music pattern interval overlap between two sequences, which is described in Algorithm \ref{alg:algorithm3}. The new metric can evaluate melodic similarity between two pieces of music if the target event is \textit{Note On}. If $PS$ is higher, then the melody of the new music is more similar to the target music, meaning that users will be more familiar with the new music. In addition, if the pattern length $p$ is longer, then the similarity will be lower because it is more difficult to find a match between two sequences. 
	
	\begin{algorithm}[tb]
		\caption{Pattern similarity.}
		\label{alg:algorithm3}
		\textbf{Input}: Music event of a user’s favorite music $X$: $\tilde{E}^{X}= \{\tilde{e}_{1}^{X},\tilde{e}_{2}^{X},\ldots,\tilde{e}_{z_{X}}^{X}\}$; music event of transferred music $\hat{Y}$: $\tilde{E}^{\hat{Y}}= \{\tilde{e}_{1}^{\hat{Y}},\tilde{e}_{2}^{\hat{Y}},\ldots,\tilde{e}_{z_{\hat{Y}}}^{\hat{Y}}\}$; music pattern length $p$
		\newline
		\textbf{Output}: Pattern similarity $PS$
		\begin{algorithmic}[1] 
			\STATE $match=0$
			\STATE $z=1$
			\STATE calculate interval of $X$ and $\hat{Y}$ and denote as $I^{X} =\{i_{1}^{X},i_{2}^{X},\ldots,i_{z_{X}-1}^{X}\} $ and $I^{\hat{Y}}=\{i_{1}^{\hat{Y}},i_{2}^{\hat{Y}},\ldots,i_{z_{\hat{Y}}-1}^{\hat{Y}}\} $ 
			\WHILE{$z<z_{\hat{Y}}-1-p$}
			\IF {$[i_{z}^{\hat{Y}},i_{z+1}^{\hat{Y}}\ldots, i_{z+p}^{\hat{Y}}]\in I^{X}$}
			\STATE  $match=match+1$
			\ENDIF
			\STATE $z=z+1$
			\ENDWHILE \newline
			\STATE $PS = \dfrac{match}{z_{\hat{Y}}-p}$ \newline
			\STATE \textbf{return} $PS$
		\end{algorithmic}
	\end{algorithm}
	\subsection{Experimental Results and Analysis}

	\subsubsection{Quantitative comparison}
	Table \ref{tab4} and \ref{tab5} list the results of our quantitative comparison. In Table \ref{tab4}, UP-Transformer achieves a significantly better average overlapped area with the user’s favorite music in terms of $D_{P}$, $D_{N}$, and $D_{D}$. Meanwhile, $D_{IOI}$ is highly competitive, implying that the new music transferred by UP-Transformer is most similar to the user’s favorite music compared with other models. Music transferred by other Transformer-based models (PopMT\_T and PopMAG\_T) is less similar than the target music compared to the input music because no prior music knowledge is included in the model. CycleGAN demonstrates unsatisfactory performance due to limited training samples. TTM uses the matrix representation method and learns music features jointly from two types of input; it does not learn event information specifically, including \textit{Note On}.
	
	In Table \ref{tab5}, UP-Transformer also leads to the highest average overlapped area with the input music in terms of $D_{N}$, $D_{D}$, and $D_{IOI}$ without losing substantial \textit{Note On} information from the input music. In particular, $D_{D}$ and $D_{IOI}$ of PopMT\_T and PopMAG\_T are also high because only \textit{Note On} events are transferred in these models. Furthermore, CycleGAN achieves poor performance given that training a large number of model parameters based on only two music samples is challenging.

	\subsubsection{Qualitative comparison}
	We also conducted a qualitative experiment because the best way to evaluate transferred music is by listening. We distributed a listening test to 23 subjects who have been trained to play a musical instrument and were familiar with basic music theory. Subjects were asked to provide their favorite music before the experiment. In the experiment, each subject was first asked to listen to two music samples: a piece of the original input music and a piece of their favorite music. Each subject was next asked to listen to five pieces of randomly chosen music that were transferred by PopMT, PopMAG, CycleGAN, TTM, and UP-Transformer. All subjects were then asked to evaluate the following three metrics after listening to each piece of music on a 5-point scale (1 is lowest, 5 is highest). 
	
	$\bullet$ Musicality: How do you perceive the musicality of the music?
	
	$\bullet$ Similarity: How similar is the music to your favorite music?
	
	$\bullet$ Preference: How much do you like the music?

	\begin{table}
		\centering
		\setlength\tabcolsep{6pt}
		\caption{Model comparison. The ground truth is the user's favorite music.  $D_{P}, D_{N}, D_{D}, D_{IOI}$: average overlapped area of pitch class, note, duration, and inter-onset interval.}
		\begin{tabular}{ccccc}
			\toprule
			Music & $D_{P}$&$D_{N}$& $D_{D}$& $D_{IOI}$ \\
			\midrule
			Input &0.60$\pm$.04& 0.24$\pm$.05& 0.34$\pm$.11& 0.25$\pm$.08\\
			PopMT\_T \cite{huang2020pop} &0.58$\pm$.10&0.21$\pm$.06&	0.33$\pm$.20&\textbf{0.27$\pm$.15}\\
			PopMAG\_T \cite{ren2020popmag} &0.60$\pm$.12&0.23$\pm$.05&0.33$\pm$.20&0.26$\pm$.15\\
			CycleGAN \cite{brunner2018symbolic} &0.27$\pm$.15&0.13$\pm$.07&0.08$\pm$.06&0.04$\pm$.03\\
			TTM \cite{hu2020make} &0.57$\pm$.10&0.29$\pm$.12&0.27$\pm$.17&0.14$\pm$.10\\
			UP-Transformer &\textbf{0.86$\pm$.10}&\textbf{0.70$\pm$.13}&\textbf{0.34$\pm$.11}&0.26$\pm$.08\\
			\bottomrule
		\end{tabular}
		\label{tab4}
	\end{table}
	\begin{table}
		\setlength\tabcolsep{6pt}
		\centering
		\caption{Model comparison. The ground truth is the input music.}
		\begin{tabular}{ccccc}
			\toprule
			Music & $D_{P}$&$D_{N}$& $D_{D}$& $D_{IOI}$ \\
			\midrule
			PopMT\_T \cite{huang2020pop} &\textbf{0.59$\pm$.15}&0.20$\pm$.07&0.85$\pm$.07&0.71$\pm$.06\\
			PopMAG\_T \cite{ren2020popmag} &0.53$\pm$.10&0.21$\pm$.04&0.86$\pm$.10&0.71$\pm$.08\\
			CycleGAN \cite{brunner2018symbolic} &0.26$\pm$.15&0.12$\pm$.08&0.15$\pm$.14&0.05$\pm$.05\\
			TTM \cite{hu2020make} &0.55$\pm$.05&0.21$\pm$.05&0.25$\pm$.03&0.15$\pm$.03\\
			UP-Transformer &0.57$\pm$.05&\textbf{0.22$\pm$.04}&\textbf{0.91$\pm$.04}&\textbf{0.78$\pm$.04}\\
			\bottomrule
		\end{tabular}
		\label{tab5}
	\end{table}
	\begin{figure}[!t]

		{\includegraphics[width=\linewidth,trim=80 10 110 43,clip]{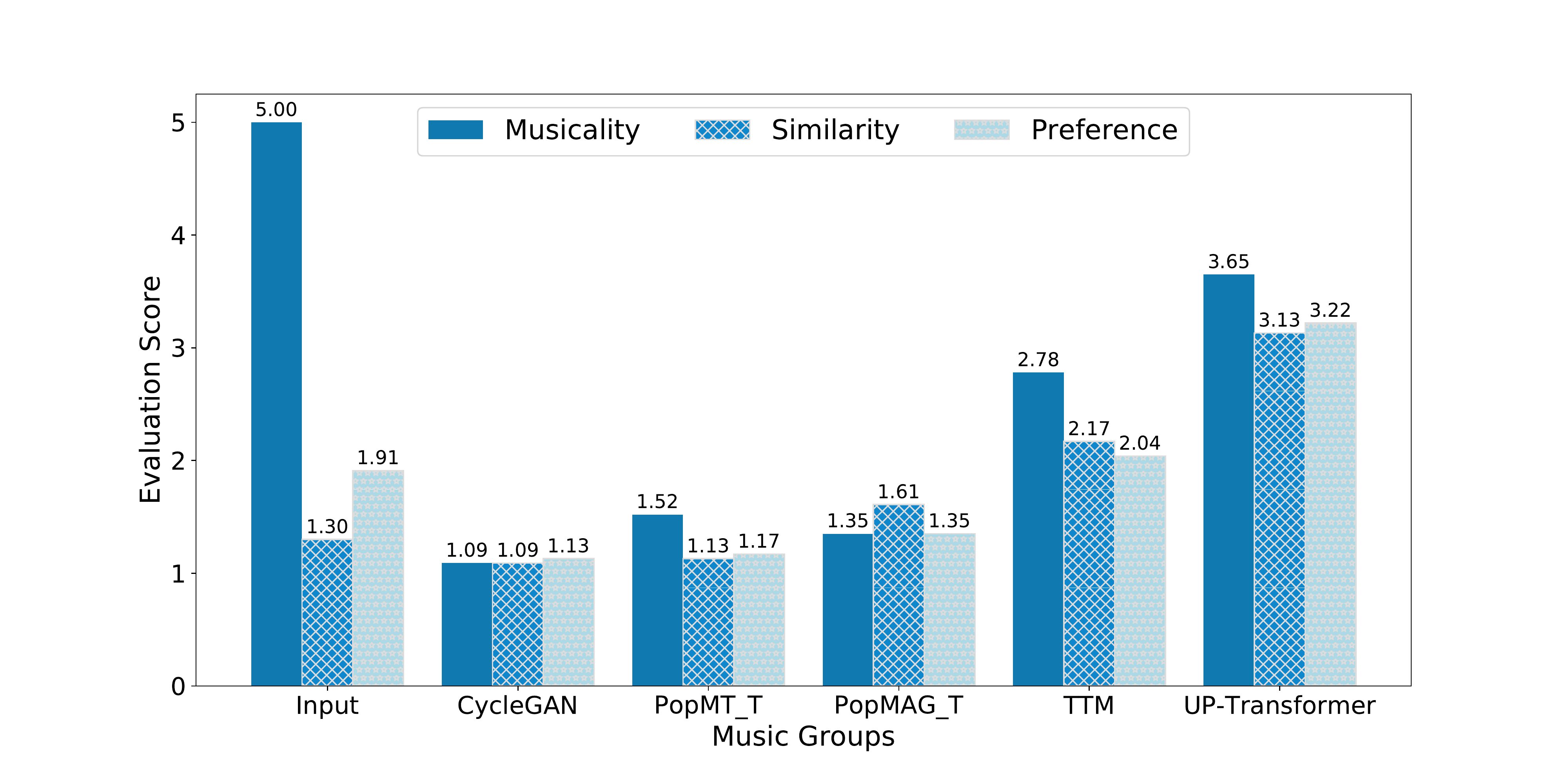}}
		\caption{Qualitative experiments’ performance on six groups of music.}
		\label{fig5}
		
	\end{figure}
	Fig. \ref{fig5} depicts the qualitative results of the input music, CycleGAN, PopMT\_T, PopMAG\_T, MMT, and UP-Transformer, results are summarized as follows:  
	
	$\bullet$ In terms of musicality, although a gap is apparent between human-composed and transferred musical pieces, UP-Transformer can transfer music with the highest quality compared with other models.  
	
	$\bullet$ In terms of similarity, UP-Transformer also achieves the highest performance. Compared with the input music, the similarity score rises from 1.30 to 3.13, suggesting that music transferred by UP-Transformer is more similar to the user's favorite music compared to the input music. This result implies that some music patterns from the user's favorite music are learned from UP-Transformer. 
	
	$\bullet$ In terms of preference, the score increases from 1.91 to 3.22, meaning that UP-Transformer can transfer input music to be similar to the user's preference. Compared with other models, UP-Transformer again exhibits the best performance. Users should therefore be more satisfied with music transferred by UP-Transformer. 
	
	\subsection{Ablation Study}
		In this subsection, we first conduct two experiments without implementing two steps to determine 1) the best music event to be transferred and 2) the most appropriate sequence length for transfer. Then, we perform three experiments to validate 3) the effect of using Transformer-XL as the backbone model, 4) the effect of using the favorite-aware loss function in the fine-tuning phase, and 5) the effect of implementing two steps in the transfer phase. 
	\subsubsection{Transferring different events}
	\begin{figure*}
		\hspace*{\fill}%
		\subfigure[original music] {\includegraphics[width=1.66in,height=1.2in,trim=60 35 40 43,clip]{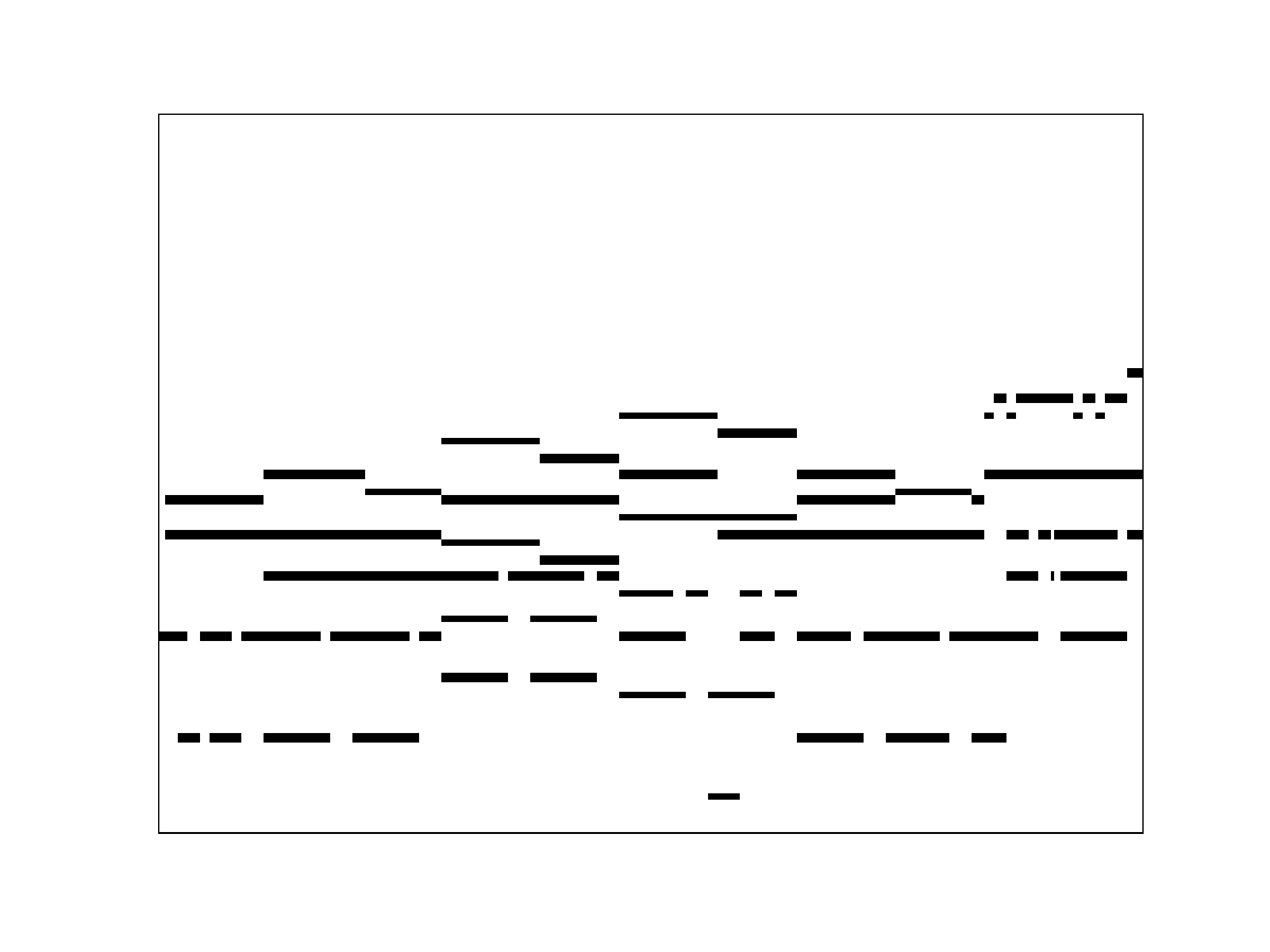}}
		\subfigure[transferring \textit{NO}] {\includegraphics[width=1.66in,height=1.2in,trim=60 35 40 43,clip]{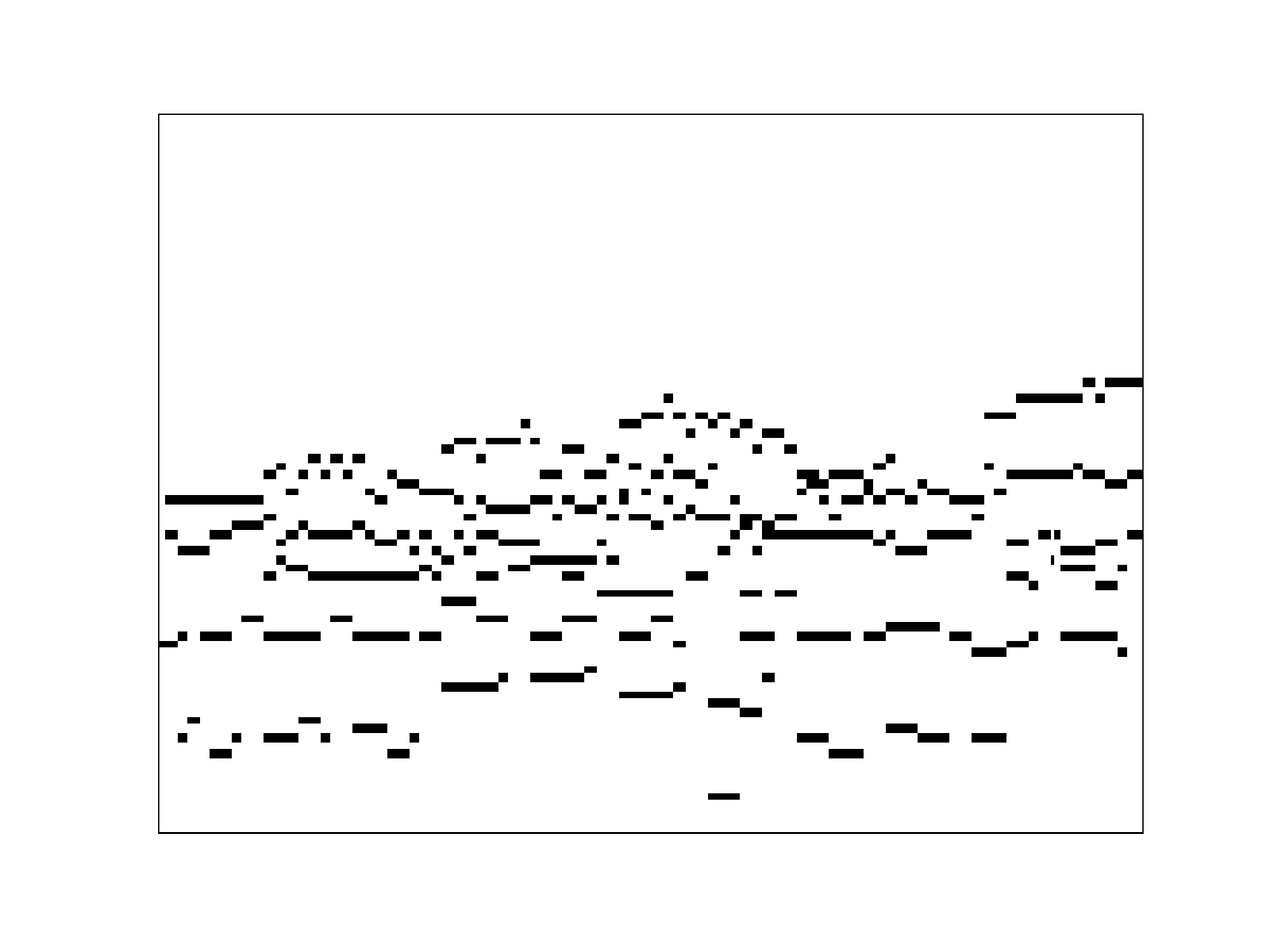}}
		\subfigure[transferring \textit{ND}] {\includegraphics[width=1.66in,height=1.2in,trim=60 35 40 43,clip]{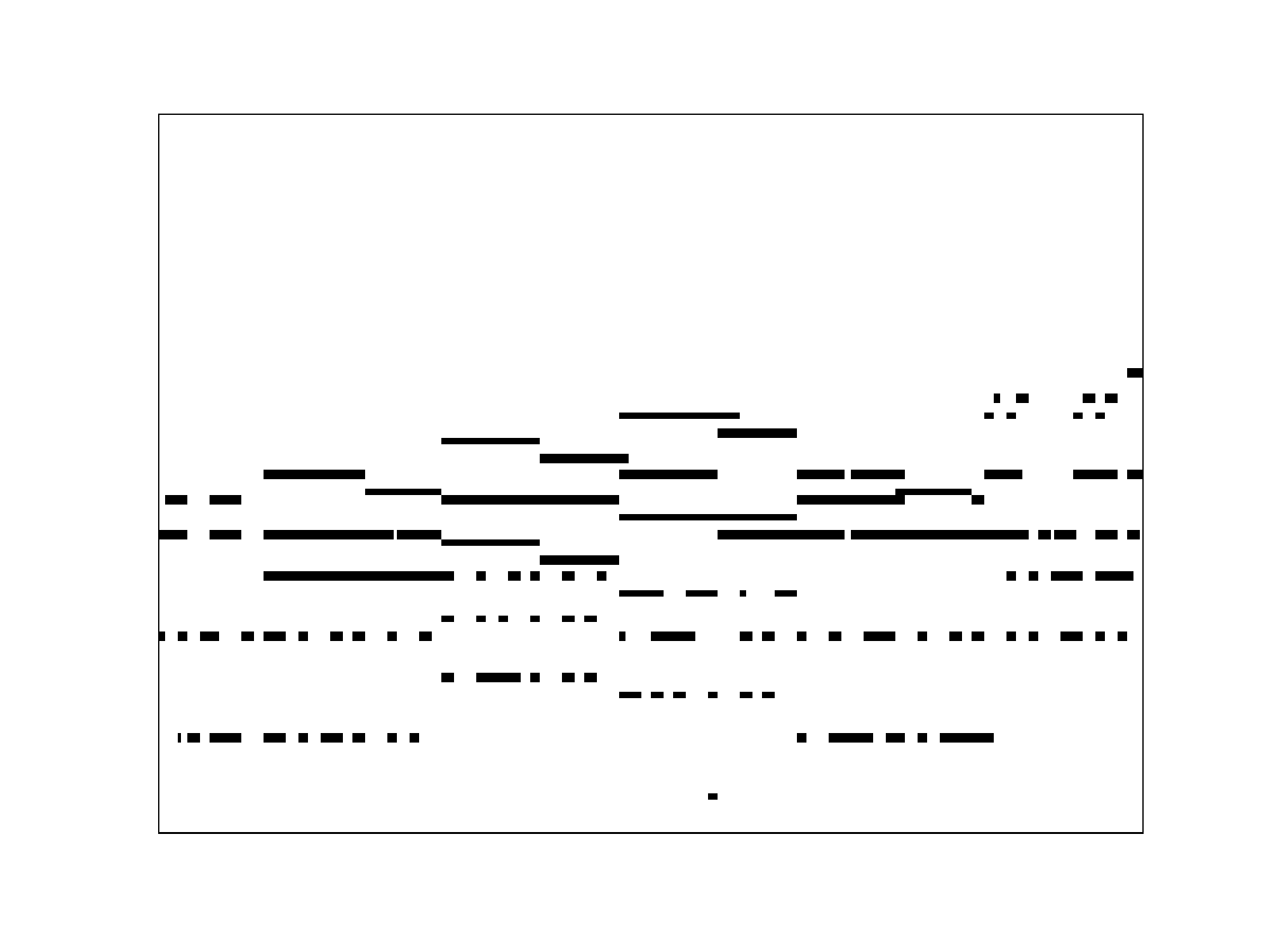}}
		\subfigure[transferring \textit{P}] {\includegraphics[width=1.66in,height=1.2in,trim=60 35 40 43,clip]{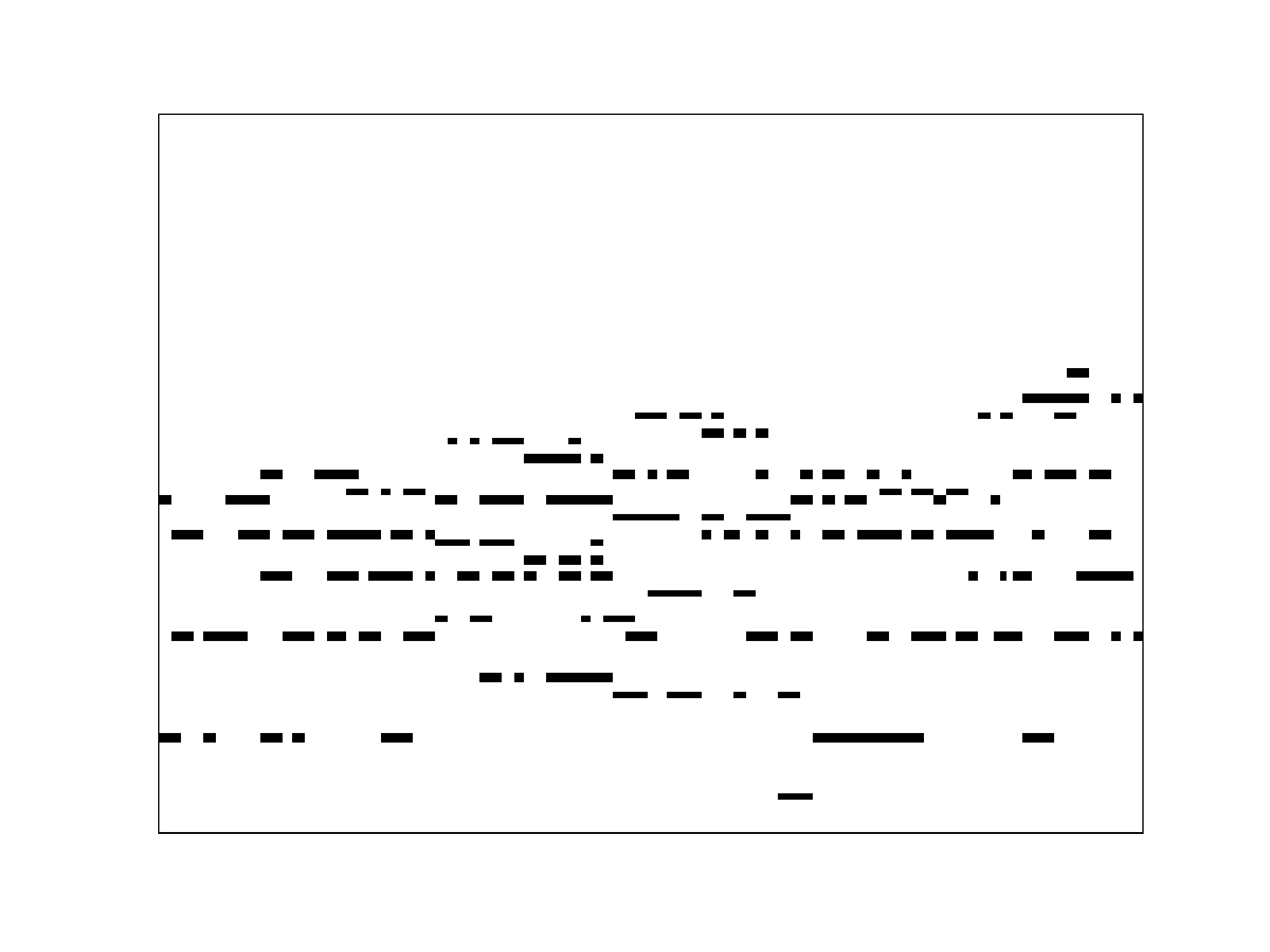}}
		\caption{One example of transferring different events.}
		\label{fig7}
	\end{figure*}
	
	We first performed an experiment to explore music performance when transferring different events. Based on the mean value in Table \ref{tab1}, \textit{Note On}, \textit{Note Duration}, and \textit{Position} appear more frequently than other events because they constitute notes and exert more significant influences on the output music. As such, we compared performance when transferring one of these three events. We did not transfer \textit{Note Velocity} because transferring velocity is trivial in this work; it only changes the loudness of music, which does not influence the music’s structure and melody.
	
	To verify the transfer effects of different music events, we first illustrate the results of transferring \textit{Note On}, \textit{Note Duration}, and \textit{Position}, as shown in Fig. \ref{fig7}. Four examples represent original music samples and samples after transferring \textit{Note On}, \textit{Note Duration}, and \textit{Position}, respectively. Music tends to be different when transferring these events because the latter three samples vary from the first sample, but transferring \textit{Note On} has a more significant impact on music melodies: the music patterns are different in Fig. \ref{fig7} (b), resulting in greater diversity.
	
	Next, we randomly chose one piece of input music for each user to realize transfer. Transferred results are compared based on four metrics, $D_{P}$, $D_{N}$, $D_{D}$, and $D_{IOI}$, respectively. Table \ref{tab111} shows the performance when $\tilde{E}=e_{P}$, $\tilde{E}=e_{ND}$, and $\tilde{E}=e_{NO}$. When comparing all music (including the input music) with the user’s favorite music, the following findings are observed:
	\begin{enumerate}
		\item The differences in four measures are negligible when transferring \textit{Position}. This finding is reasonable given that \textit{Position} does not directly relate to pitch value or duration value. However, changing \textit{Position} will indirectly affect the note duration and inter-onset interval as the position of each note is changed, leading to fluctuations in the result. 
		
		\item Transferring \textit{Note Duration} leads to a higher value of both $D_{D}$ and $D_{IOI}$, but $D_{P}$ and $D_{N}$ do not change because these metrics measure note-on information.

		\item Transferring \textit{Note On} results in a higher value of $D_{P}$ and $D_{N}$. It is also worth mentioning that $D_{D}$ and $D_{IOI}$ might change even though only \textit{Note On} events are transferred, as pitch value will influence the duration of consecutive notes. For example, if two consecutive notes transfer from different pitch values to the same pitch value, then the note duration and inter-onset interval will change. 
	\end{enumerate}
	\begin{table}
		\setlength\tabcolsep{9pt}
		\centering
		\caption{Performance when transferring different music events. The ground truth is the user’s favorite music. $D_{P}, D_{N}, D_{D}, D_{IOI}$: average overlapped area of pitch class, note, duration, and inter-onset interval. \textit{P}, \textit{ND}, \textit{NO}: \textit{Position}, \textit{Note Duration}, \textit{Note On}. }
		\begin{tabular}{ccccc}
			\toprule
			Music &$D_{P}$&$D_{N}$& $D_{D}$& $D_{IOI}$ \\
			\midrule
			Input &0.62$\pm$.12& 0.24$\pm$.07&0.14$\pm$.07&0.21$\pm$.09\\
			Transfer \textit{P} &0.62$\pm$.12& 0.24$\pm$.07&0.18$\pm$.12&0.18$\pm$.09\\
			Transfer \textit{ND}&0.62$\pm$.12& 0.24$\pm$.07& \textbf{0.60$\pm$.16}&\textbf{0.24$\pm$.07}\\ 	
			Transfer \textit{NO}& \textbf{0.75$\pm$.10}&\textbf{0.39$\pm$.04}&0.14$\pm$.07&0.22$\pm$.08\\
			\bottomrule
		\end{tabular}
		\label{tab111}
	\end{table}
	
	Overall, transferring \textit{Note On} and \textit{Note Duration} can lead to improved output music. 
	To measure melodic similarity and further listen to the music samples, in the following experiment, we set $\tilde{E} = e_{NO}$; that is, \textit{Note On} events were transferred. 
	
	\subsubsection{Transferring on different music sequence lengths}
	
	\begin{figure}[!t]

		{\includegraphics[width=\linewidth,trim=70 20 100 50,clip]{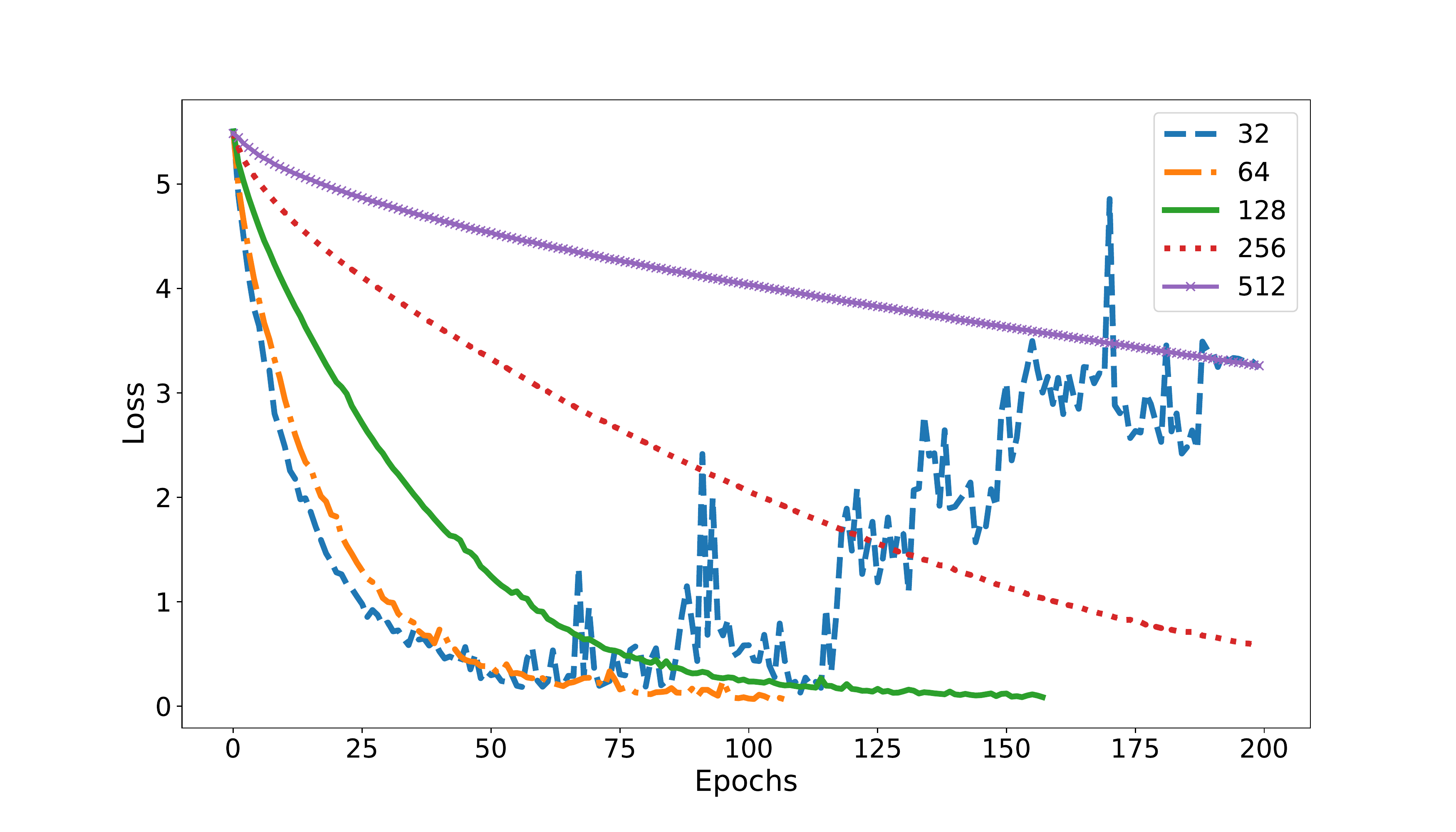}}
		\caption{Fine-tuning performance with different music sequence lengths.}
		\label{fig6}
		
	\end{figure}
	
	\begin{table}
		\setlength\tabcolsep{5pt}
		\centering
		\caption{UP-Transformer performance with different music sequence lengths. The ground truth is the user’s favorite music. $D_{P}, D_{N}, D_{D}, D_{IOI}$: average overlapped area of pitch class, note, duration, and inter-onset interval.}
		\begin{tabular}{ccccc}
			\toprule
			Music Sequence Length & $D_{P}$&$D_{N}$& $D_{D}$& $D_{IOI}$ \\
			\midrule
			32  &0.66$\pm$.17&0.37$\pm$.05	&0.14$\pm$.07&0.22$\pm$.08\\
			64  &0.75$\pm$.12&0.37$\pm$.05&0.14$\pm$.07&0.22$\pm$.08\\
			128  &\textbf{0.75$\pm$.10}&\textbf{0.39$\pm$.04}&\textbf{0.14$\pm$.07}&\textbf{0.22$\pm$.08}\\
			256 &0.67$\pm$.18&0.36$\pm$.05&0.14$\pm$.07&0.21$\pm$.08\\
			512 &0.68$\pm$.15&0.39$\pm$.04&0.14$\pm$.07&0.22$\pm$.08\\
			\bottomrule
		\end{tabular}
		\label{tab7}
	\end{table}
	\begin{table}
		\setlength\tabcolsep{6pt}
		\centering
		\caption{Pattern similarity with different music sequence lengths. $p$ is the music pattern length.}
		\begin{tabular}{ccccc}
			\toprule
			Music Sequence Length & p=2&p=3& p=4& p=5\\
			\midrule
			32  &60.26\%&10.00\%&1.57\%&0.47\%\\
			64  &\textbf{66.45\%}&\textbf{12.72\%}&1.61\%&0.44\%\\
			128  &65.06\%&11.70\%&\textbf{1.67\%}&\textbf{0.53\%}\\
			256  &60.13\%&10.34\%&1.37\%&0.45\%\\
			512  &63.21\%&10.69\%&1.50\%&0.50\%\\
			\bottomrule
		\end{tabular}
		\label{tab8}
	\end{table}
	
	We also trained the model on different music sequence lengths, namely $N= 32, 64, 128, 256, 512$. Here, we compare the fine-tuning performance based on different music sequence lengths. Fig. \ref{fig6} displays the training result. The convergence speed was low if the music sequence length was too long ($N=256,512$). These two models did not converge after training for 200 epochs because we only had one music sample; a longer music sequence length indicates fewer music segments for training. If the music sequence length was too short ($N = 32$), the model could not converge and demonstrated poor performance. Therefore, only models with a sequence length of 64 or 128 returned a good model loss.
	
	Next, we compared the transferred music with a user’s favorite music based on four metrics as listed in Table \ref{tab7}. Although the performance was quite similar, transferred music achieved the best performance when training the model with a music sequence length equal to 128. We further compared pattern similarity among different music sequence lengths as indicated in Table \ref{tab8}. We can conclude that if the music pattern length $p$ is short ($p=2,3$), then a shorter music sequence length can achieve better performance because the model with a short music sequence length can also memorize short music patterns. As the music pattern length becomes longer ($p=4,5$), a longer music sequence length performs better because the model can memorize more past information. Therefore, a music sequence length of 128 was chosen after comparing all results.
	
	\subsubsection{Effects of the backbone model}

\begin{table}
	\setlength\tabcolsep{8pt}
	\centering
	\caption{Performance of different backbone models. The ground truth is the user’s favorite music.}
	\begin{tabular}{ccccc}
		\toprule
		Music &$D_{P}$&$D_{N}$& $D_{D}$& $D_{IOI}$ \\
		\midrule
		Input			&0.62$\pm$.12&0.24$\pm$.07&0.14$\pm$.07&0.21$\pm$.09\\
		CNN			&0.29$\pm$.16&0.12$\pm$.06&0.07$\pm$.02&0.04$\pm$.02\\
		RNN 							
		&0.59$\pm$.09&0.24$\pm$.06&0.14$\pm$.07&0.22$\pm$.08\\
		LSTM 							
		&0.62$\pm$.10&0.25$\pm$.06&0.14$\pm$.07&0.22$\pm$.08\\	
		TF-XL   &\textbf{0.75$\pm$.10}&\textbf{0.39$\pm$.04}&\textbf{0.14$\pm$.07}&\textbf{0.22$\pm$.08}\\	 
		\bottomrule
	\end{tabular}
	\label{backbone}
\end{table}
\begin{table}
	\setlength\tabcolsep{1.5pt}
	\centering
	\caption{Performance of different attention mechanisms. The ground truth is the user’s favorite music.}
	\begin{tabular}{ccccc}
		\toprule
		Music &$D_{P}$&$D_{N}$& $D_{D}$& $D_{IOI}$ \\
		\midrule
		TF-XL w/ local attention		
		&0.69$\pm$.15&0.36$\pm$.06&0.14$\pm$.07&0.22$\pm$.08\\	
		TF-XL w/ relative local attention 				&0.71$\pm$.06&0.36$\pm$.05&0.14$\pm$.07&0.22$\pm$.08\\			TF-XL w/ relative global attention							
		&0.74$\pm$.09&0.38$\pm$.05&0.14$\pm$.07&0.22$\pm$.08\\
		
		TF-XL   &\textbf{0.75$\pm$.10}&\textbf{0.39$\pm$.04}&\textbf{0.14$\pm$.07}&\textbf{0.22$\pm$.08}\\	 
		\bottomrule
	\end{tabular}
	\label{attention}
\end{table}

	To validate the performance of our backbone model, Transformer-XL (TF-XL), and its attention mechanism, we first compared the transferred music of TF-XL, CNN, RNN, and LSTM and then compared different attention mechanisms mentioned in \cite{huang2018music} based on four metrics as listed in Tables \ref{backbone} and \ref{attention}. In Table \ref{backbone}, TF-XL achieves a significantly better average overlapped area with the user’s favorite music in terms of $D_{P}$, $D_{N}$, indicating better performance compared to other backbone models. Similar to TTM, CNN requires matrix representation as input. It therefore does not learn event information specifically, and the sparsity of the matrix representation leads to poor performance. In Table \ref{attention}, though different attention mechanisms display similar performance, the backbone TF-XL yields slightly better results. $D_{D}$ and $D_{IOI}$ remain the same in all cases because only \textit{Note On} events are transferred.

	\subsubsection{Effects of favorite-aware loss function}

\begin{figure}
		\subcapcentertrue
		\subfigure[original music]
		{\includegraphics[width=1.13in,trim=40 0 40 43,clip]{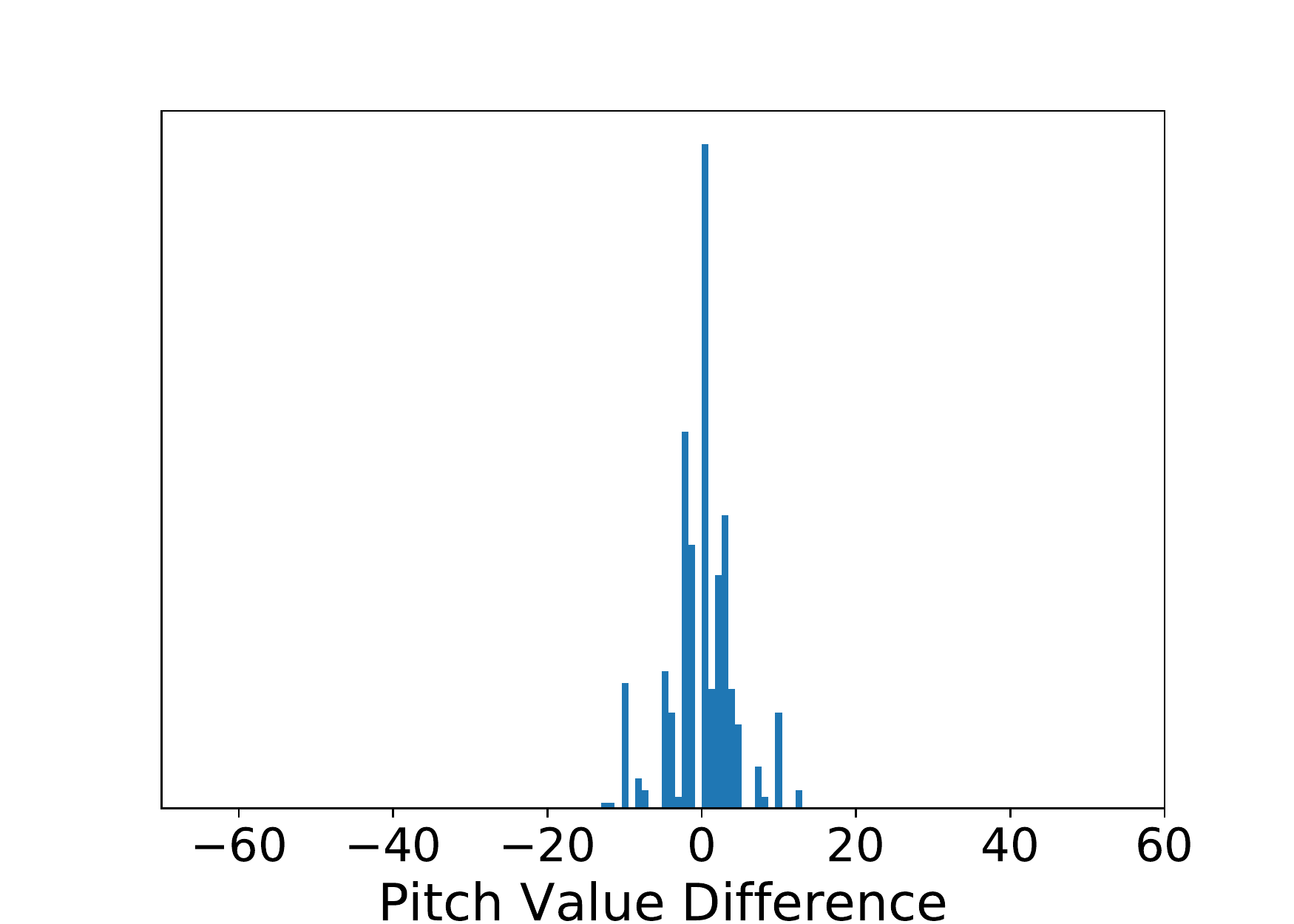}}
		\subfigure[l][music transferred without favorite-aware]
		{\includegraphics[width=1.13in,trim=40 0 40 43,clip]{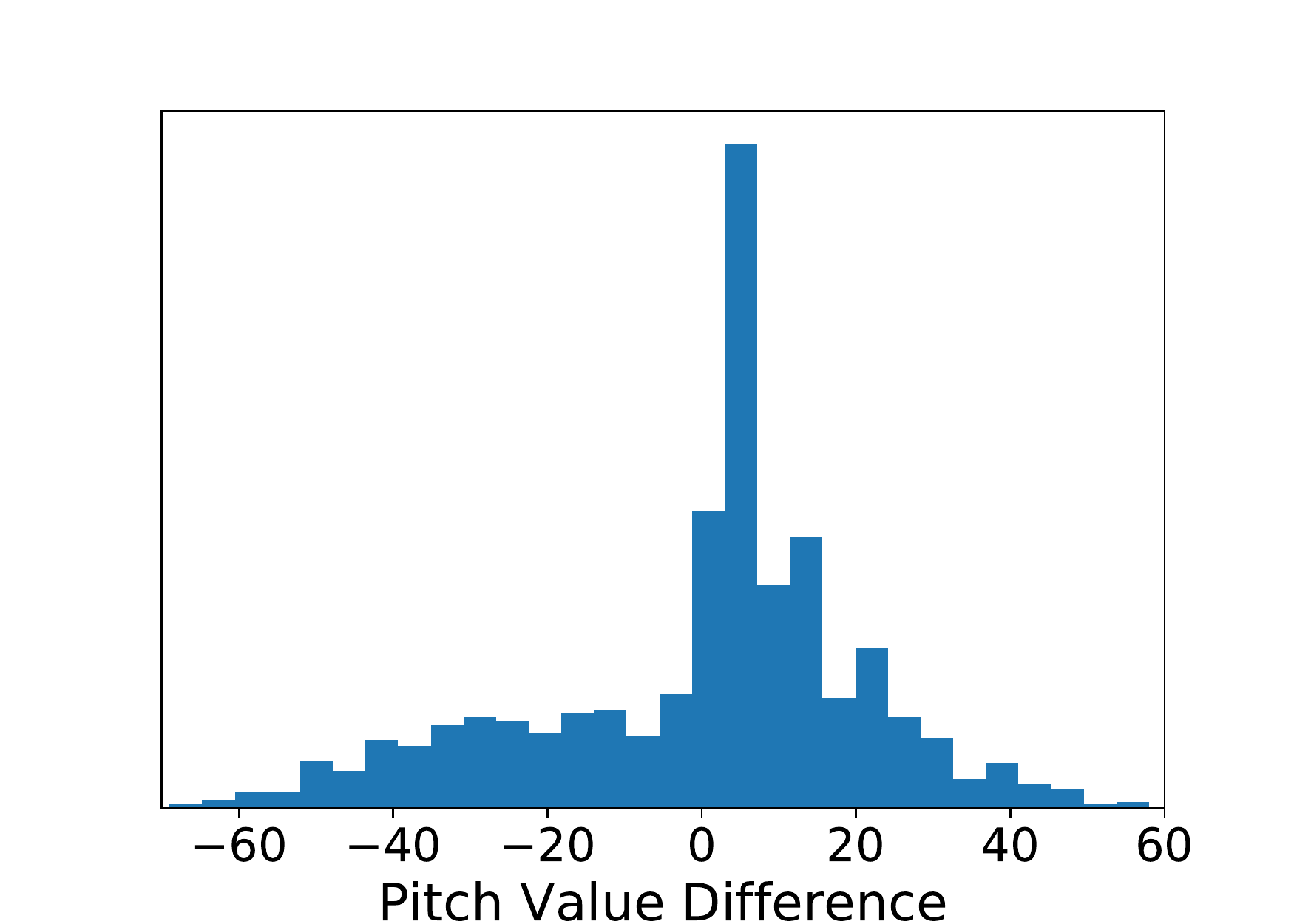}}
		\subfigure[music transferred with favorite-aware]
		{\includegraphics[width=1.13in,trim=40 0 40 43,clip]{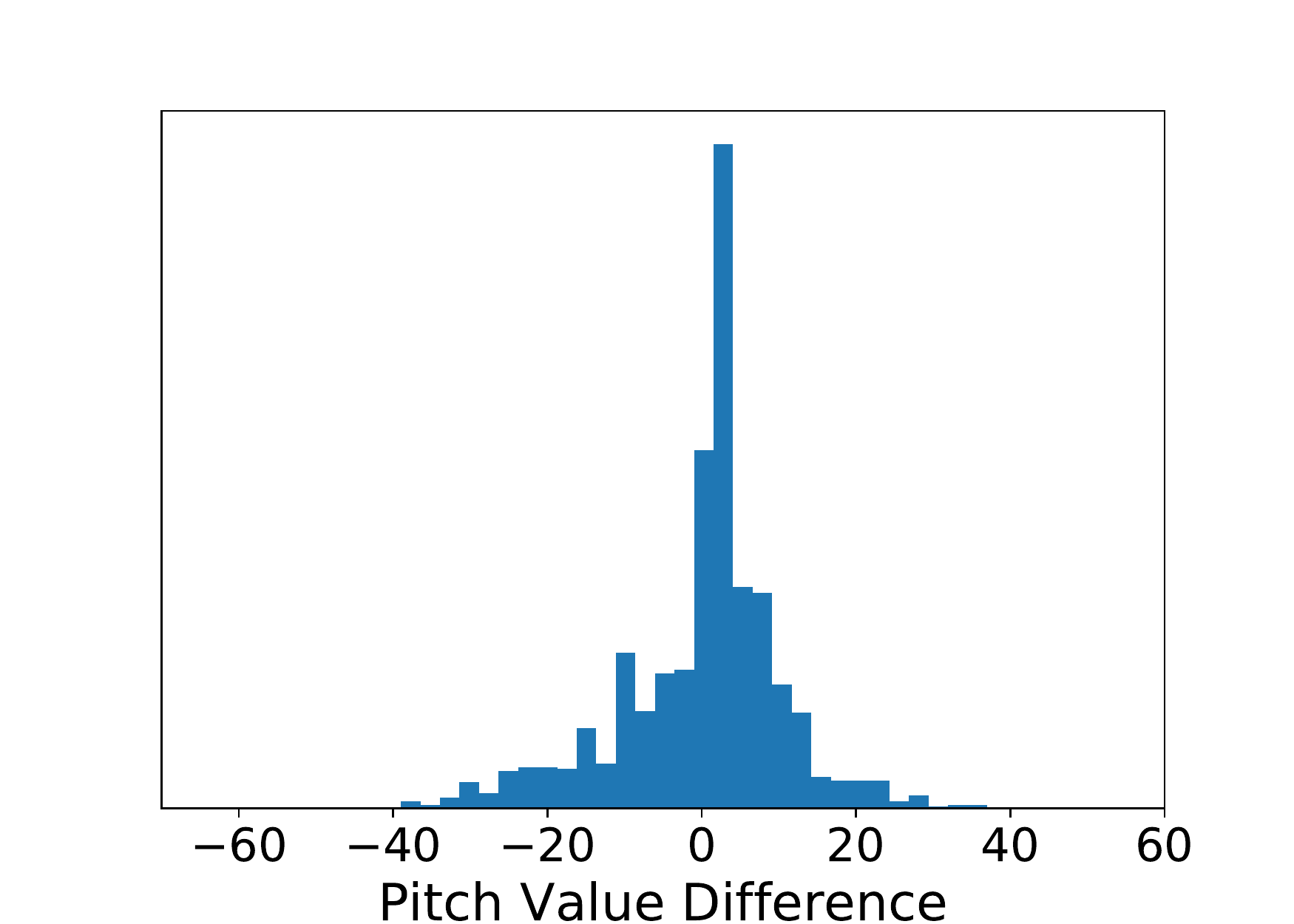}}	
		\subfigure[original music]
		{\includegraphics[width=1.13in,trim=40 0 40 43,clip]{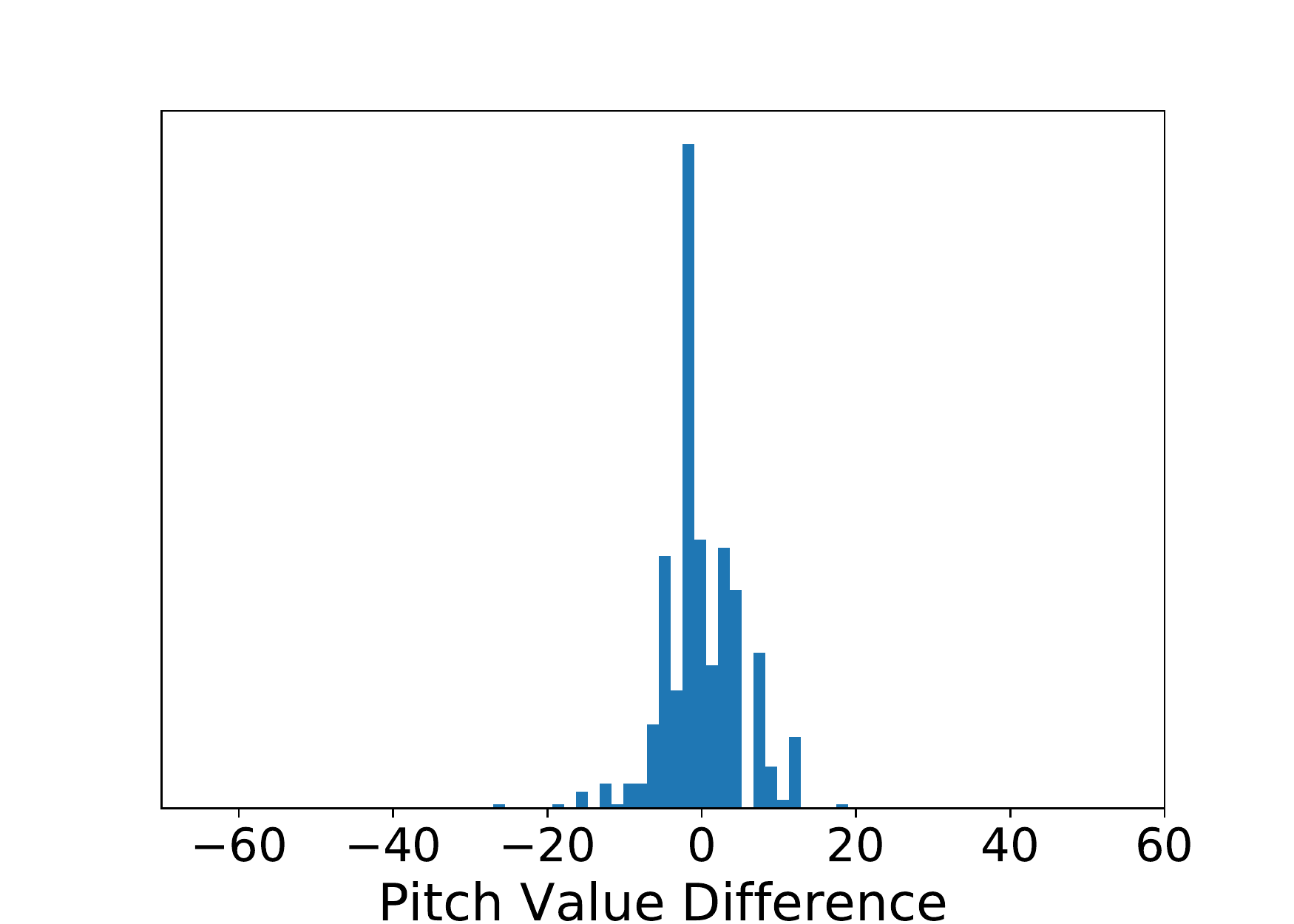}}
		\subfigure[music transferred without favorite-aware]
		{\includegraphics[width=1.13in,trim=40 0 40 43,clip]{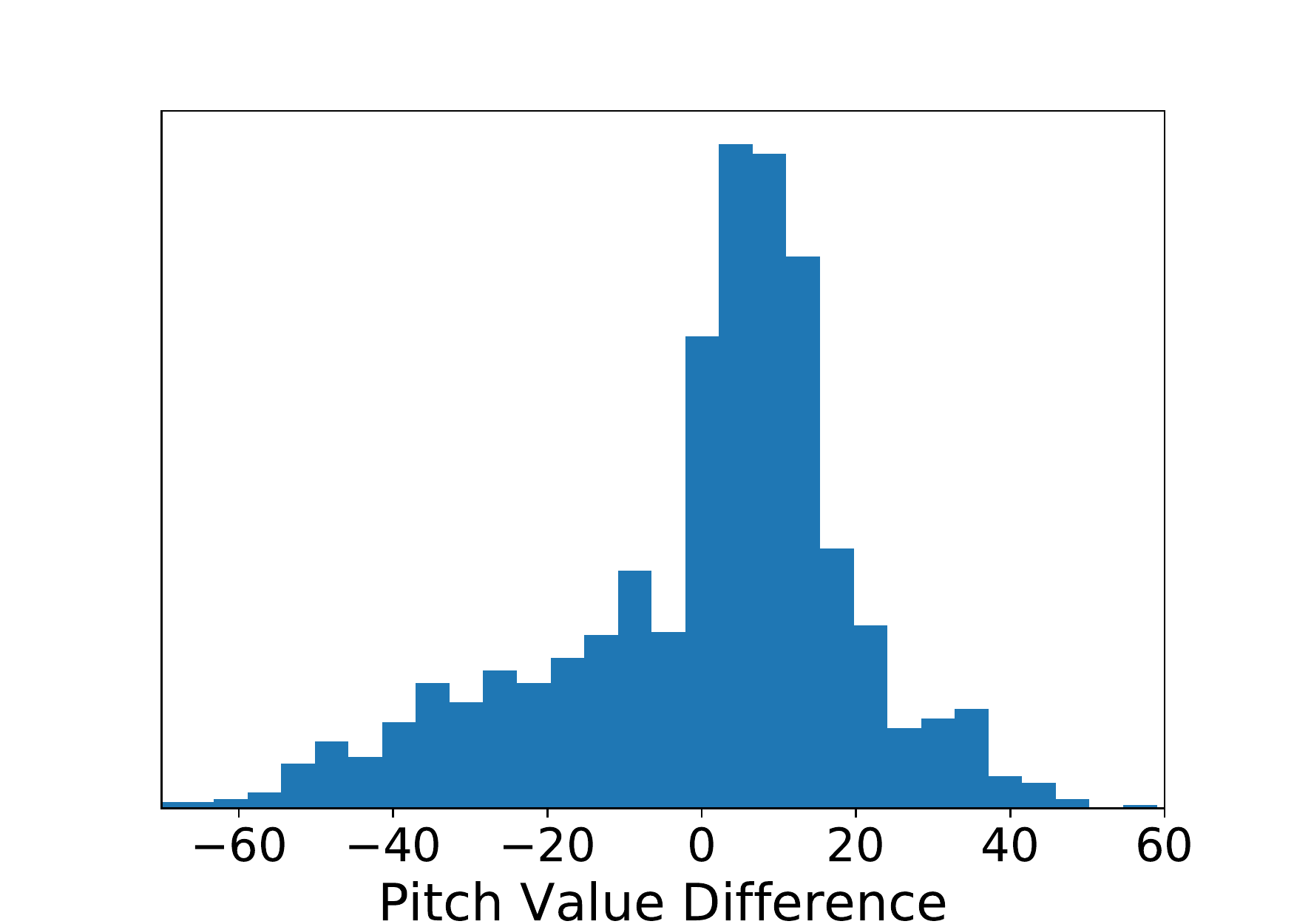}}
		\subfigure[music transferred with favorite-aware]
		{\includegraphics[width=1.13in,trim=40 0 40 43,clip]{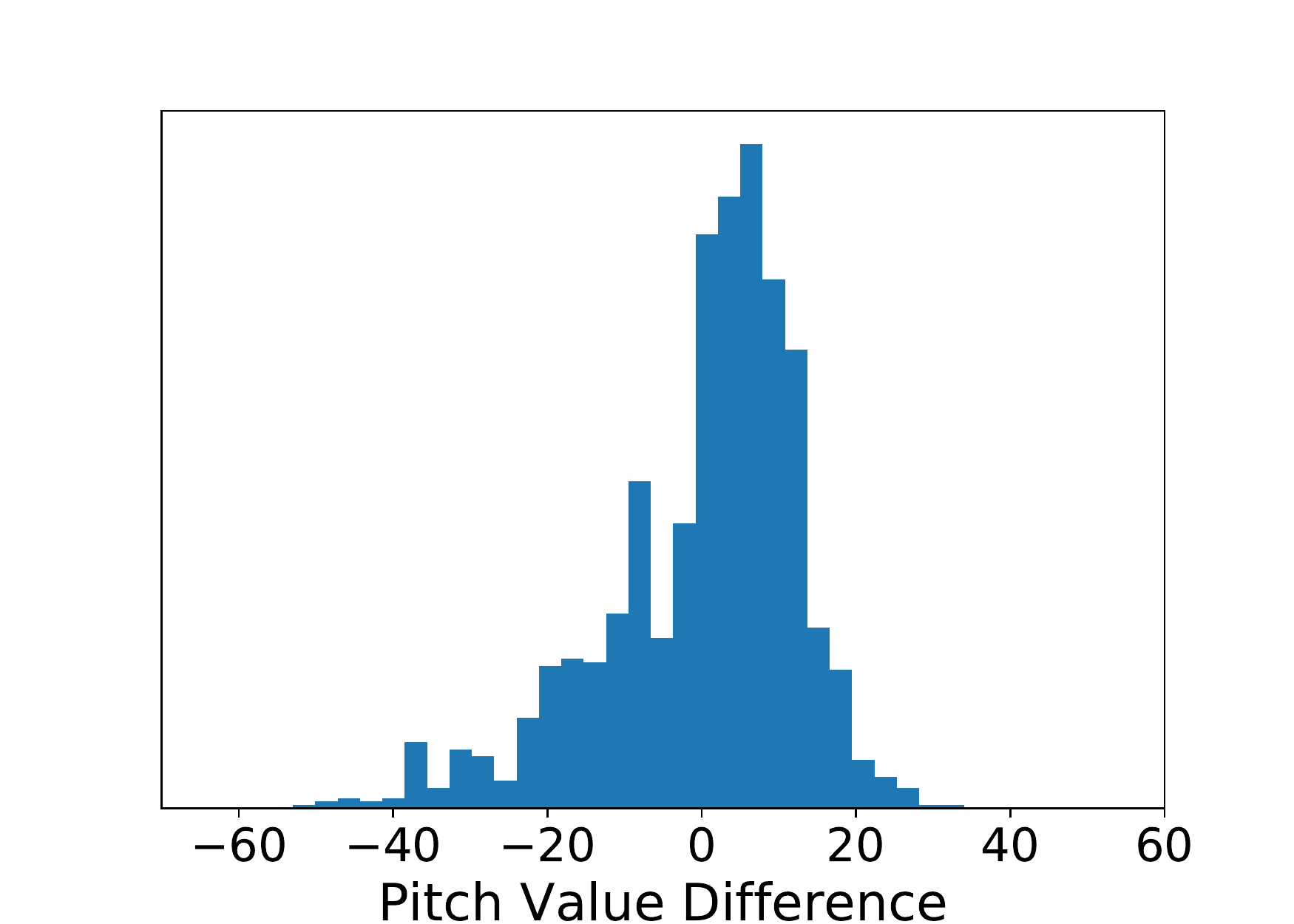}}	
		\subfigure[original music] {\includegraphics[width=1.13in,trim=40 0 40 43,clip]{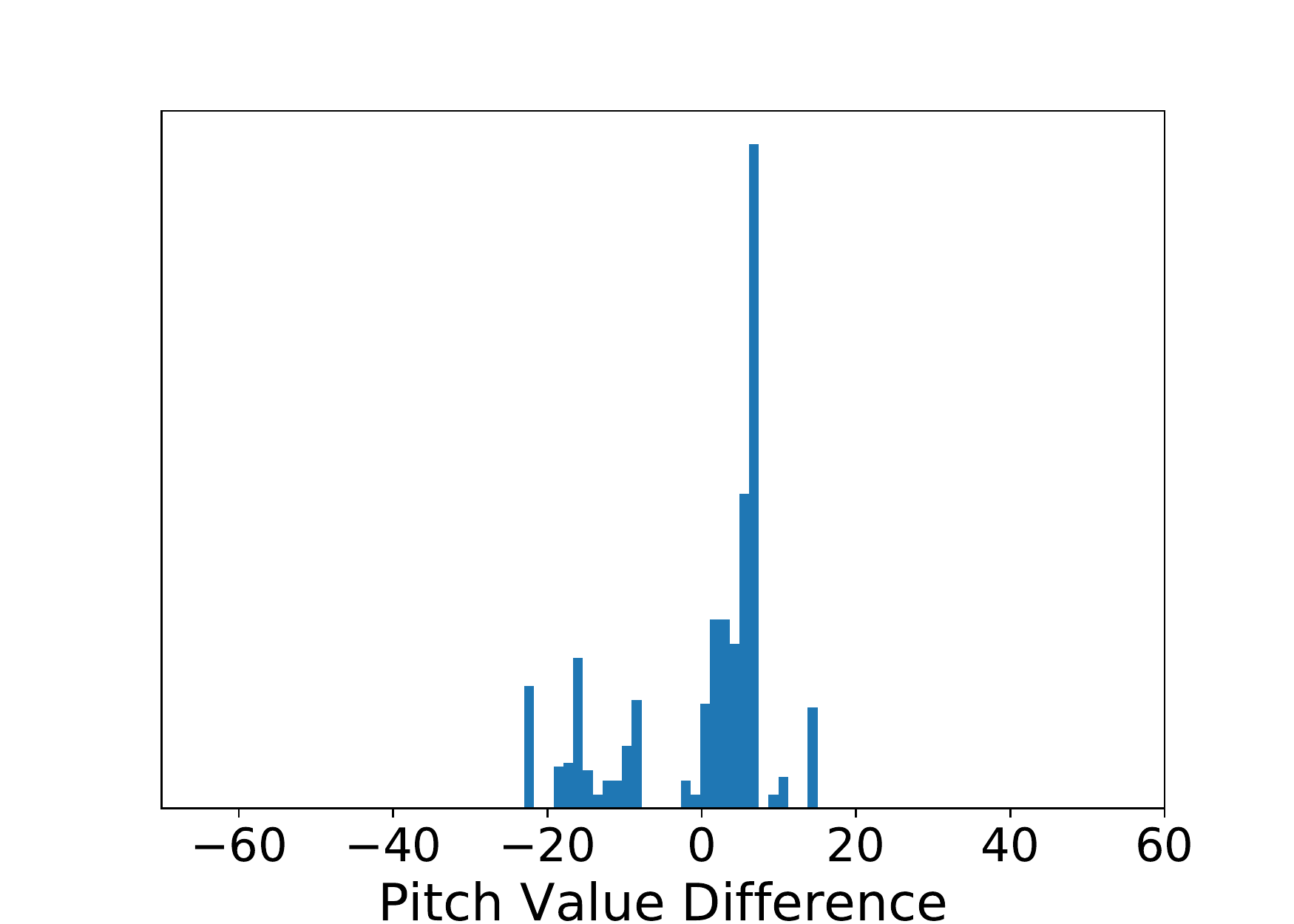}}
		\subfigure[music transferred without favorite-aware] {\includegraphics[width=1.13in,trim=40 0 40 43,clip]{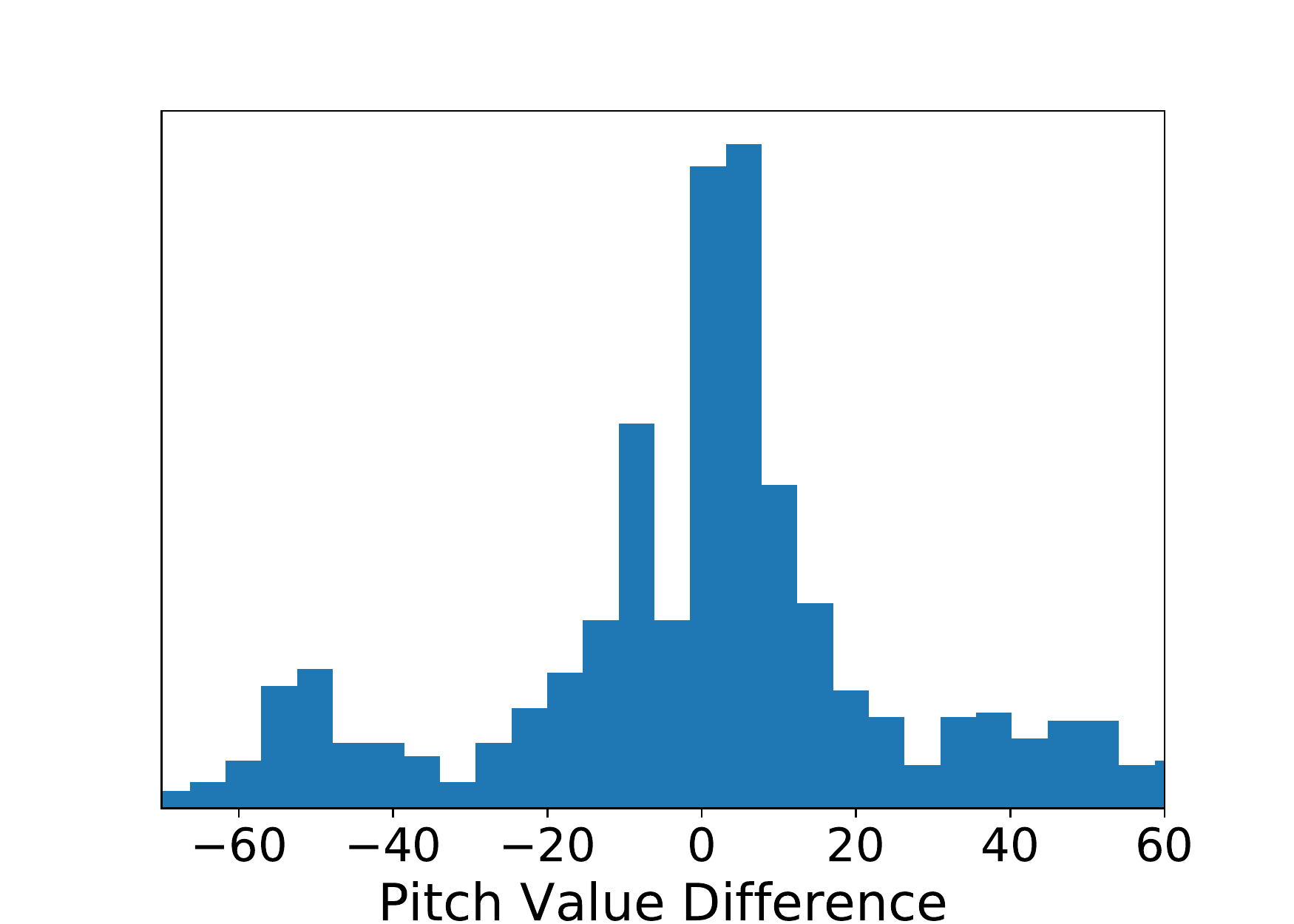}}
		\subfigure[music transferred with favorite-aware] {\includegraphics[width=1.13in,trim=40 0 40 43,clip]{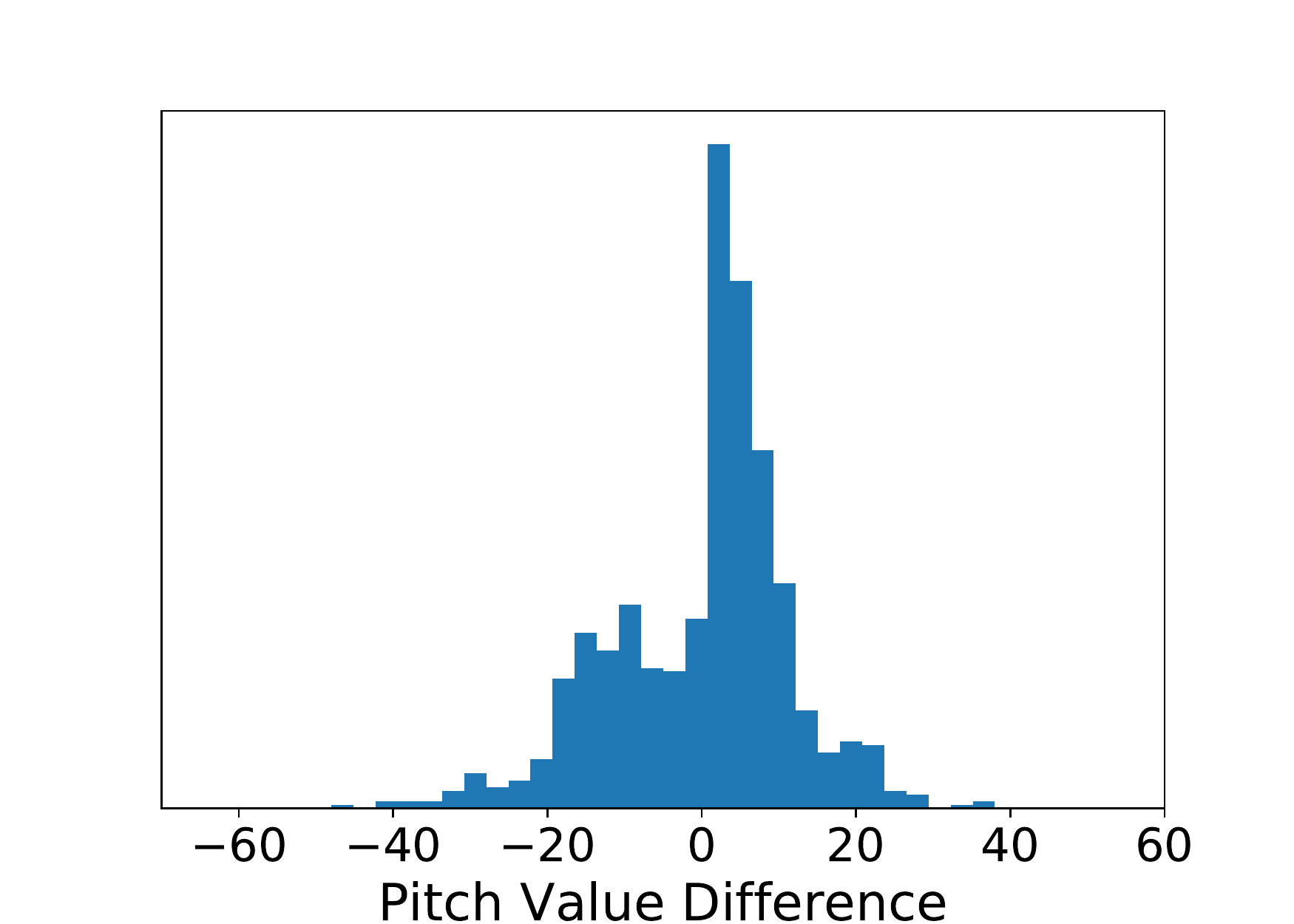}}	
		\caption{Distribution of melodic interval. Each row represents a music example.}
		\label{fig3}
		
	\end{figure}

	To validate whether the favorite-aware loss function improves fine-tuning performance, we consider the distribution of the melodic interval of the original music, music transferred without using the favorite-aware loss function, and music transferred using the favorite-aware loss function. Fig. \ref{fig3} presents three examples of these distributions. The melodic interval calculates the pitch value difference between two consecutive notes and is closely related to the music’s melody. It is assumed that music transferred using the proposed model will have a similar distribution of the melodic interval because the model should learn melodic information more precisely from the user’s favorite music after fine-tuning. Compared to the middle column of Fig. \ref{fig3}, the last column of Fig. \ref{fig3} is more similar to the melodic interval of the original music (the first column of Fig. \ref{fig3}), whereas it is obvious that the middle column of Fig. \ref{fig3} has more dramatic music interval changes because the range of the melodic interval is larger. Accordingly, the favorite-aware loss function in the fine-tuning phase helps the model learn melodic interval information from a user’s favorite music. 

	\begin{table}
		\setlength\tabcolsep{3pt}
		\centering
		\caption{UP-Transformer performance with and without implementing two steps. The ground truth is the user’s favorite music.}
		\begin{tabular}{ccccc}
			\toprule
			Music &$D_{P}$&$D_{N}$& $D_{D}$& $D_{IOI}$ \\
			\midrule
			Input &0.60$\pm$.04& 0.24$\pm$.05& 0.34$\pm$.11& 0.25$\pm$.08\\
			TF-XL w/o favorite-aware							
			&0.62$\pm$.06&0.33$\pm$.06&0.34$\pm$.11&0.26$\pm$.08\\
			TF-XL w/ favorite-aware  &0.66$\pm$.04&0.36$\pm$.06&0.34$\pm$.11&0.25$\pm$.08\\		UP-Transformer&\textbf{0.86$\pm$.10}&\textbf{0.70$\pm$.13}&\textbf{0.34$\pm$.11}&\textbf{0.26$\pm$.07}\\	 
			\bottomrule
		\end{tabular}
		\label{tab21}
	\end{table}
	In addition, Table \ref{tab21} validates the process of the favorite-aware loss function. Four sets of music are compared with the user’s favorite music in this case: input music (seven pieces of arbitrarily chosen music); music transferred without implementing the favorite-aware loss function and two steps (i.e., only implementing the cross-entropy loss function [TF-XL without favorite-aware]); music transferred without implementing two steps (TF-XL with favorite-aware); and music transferred when implementing the favorite-aware loss function and two steps (UP-Transformer). Means and standard deviations are reported. According to the first three rows, both TF-XL without favorite-aware and TF-XL with favorite-aware can generate events that are more similar to the user’s favorite music because the average overlapped area between the generated music and the user’s favorite music increases in terms of $D_{N}$ and $D_{P}$. Accordingly, TF-XL with favorite-aware benefits from the favorite-aware loss function and learns music events from the user’s favorite music more precisely, such that the generated music is more similar to the user’s favorite music.

	\subsubsection{Effects of two steps}

	\begin{figure*}
		\subfigure[p = 2]
		{\includegraphics[width=1.7in,trim=10 0 40 50,clip]{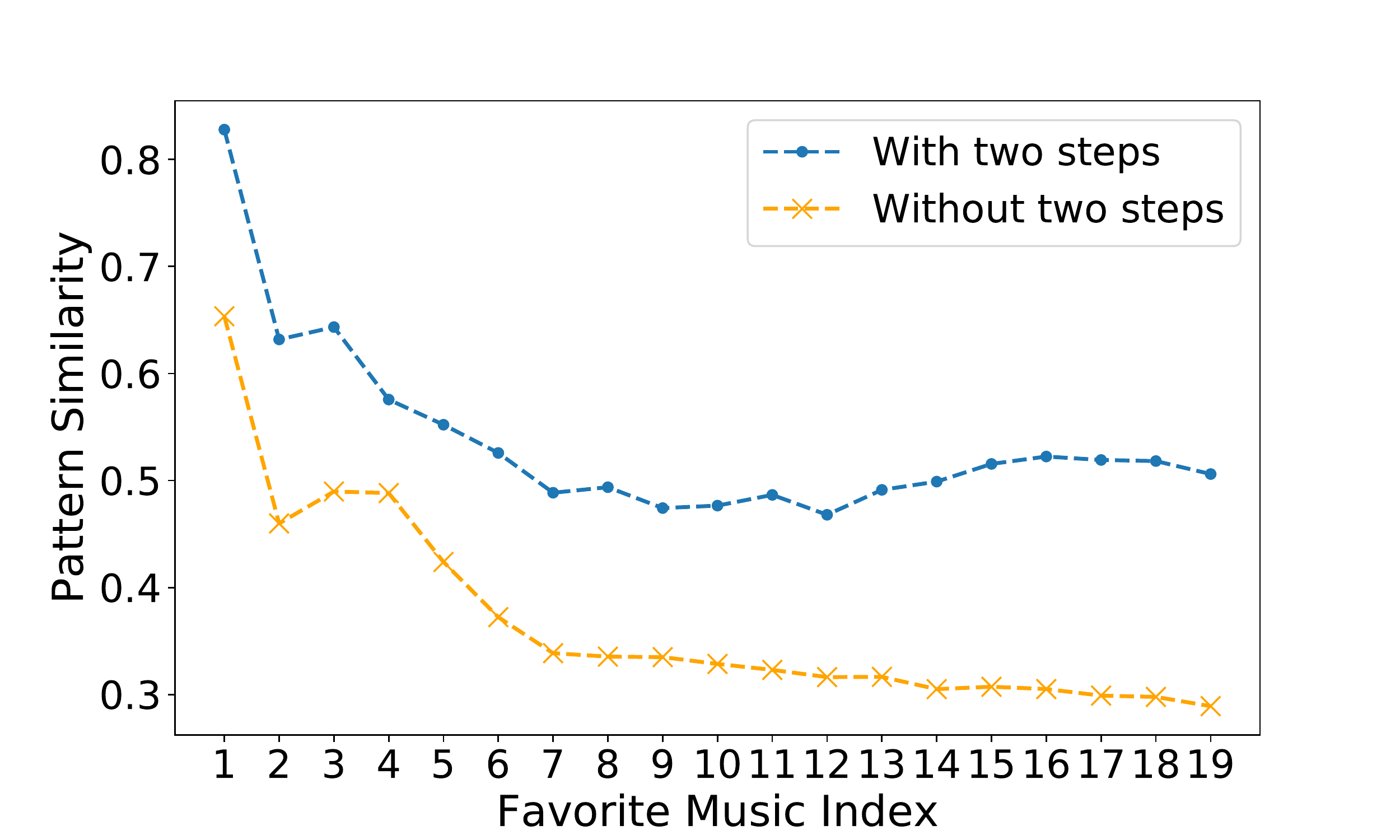}}
		\subfigure[p = 3]
		{\includegraphics[width=1.7in,trim=10 0 40 50,clip]{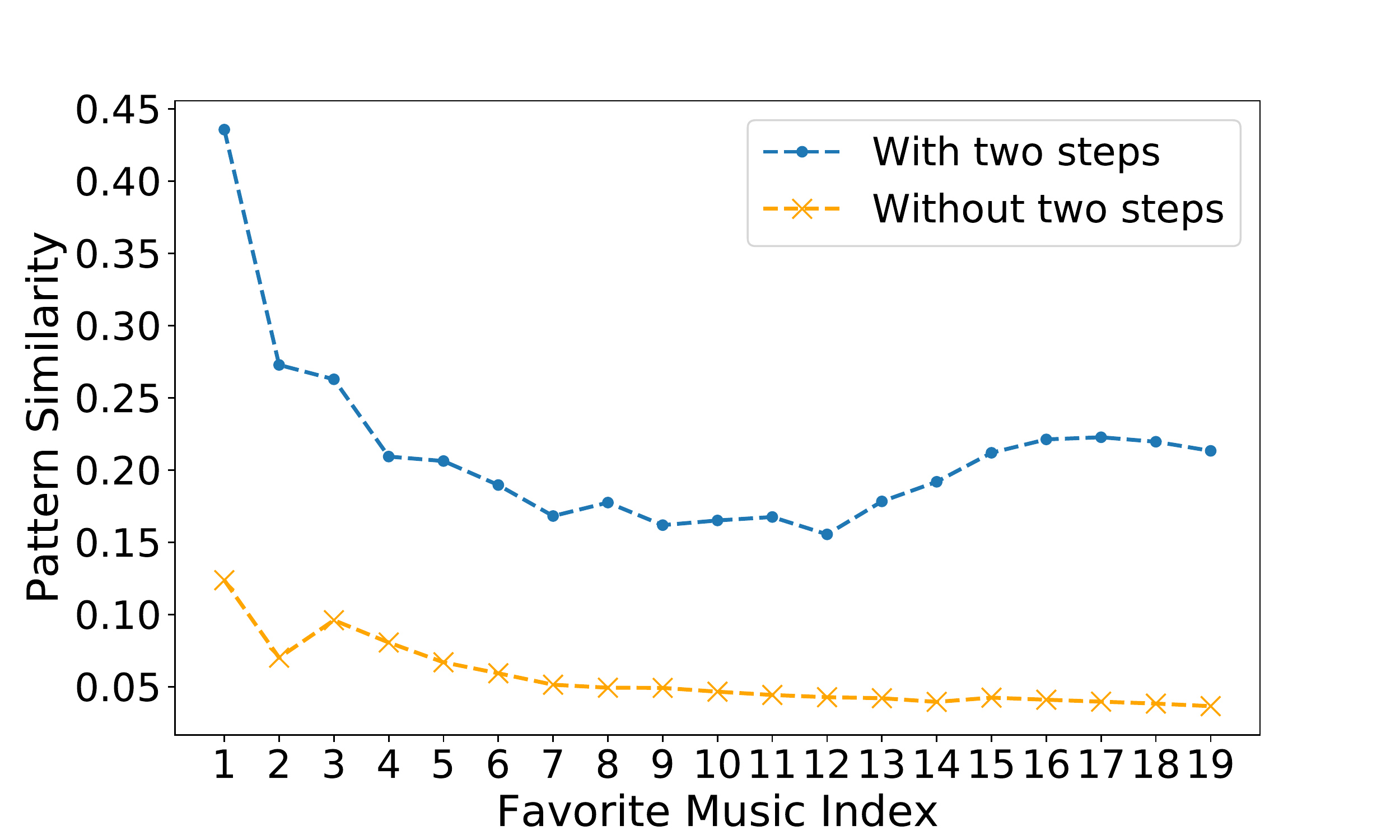}}
		\subfigure[p = 4]
		{\includegraphics[width=1.7in,trim=10 0 40 50,clip]{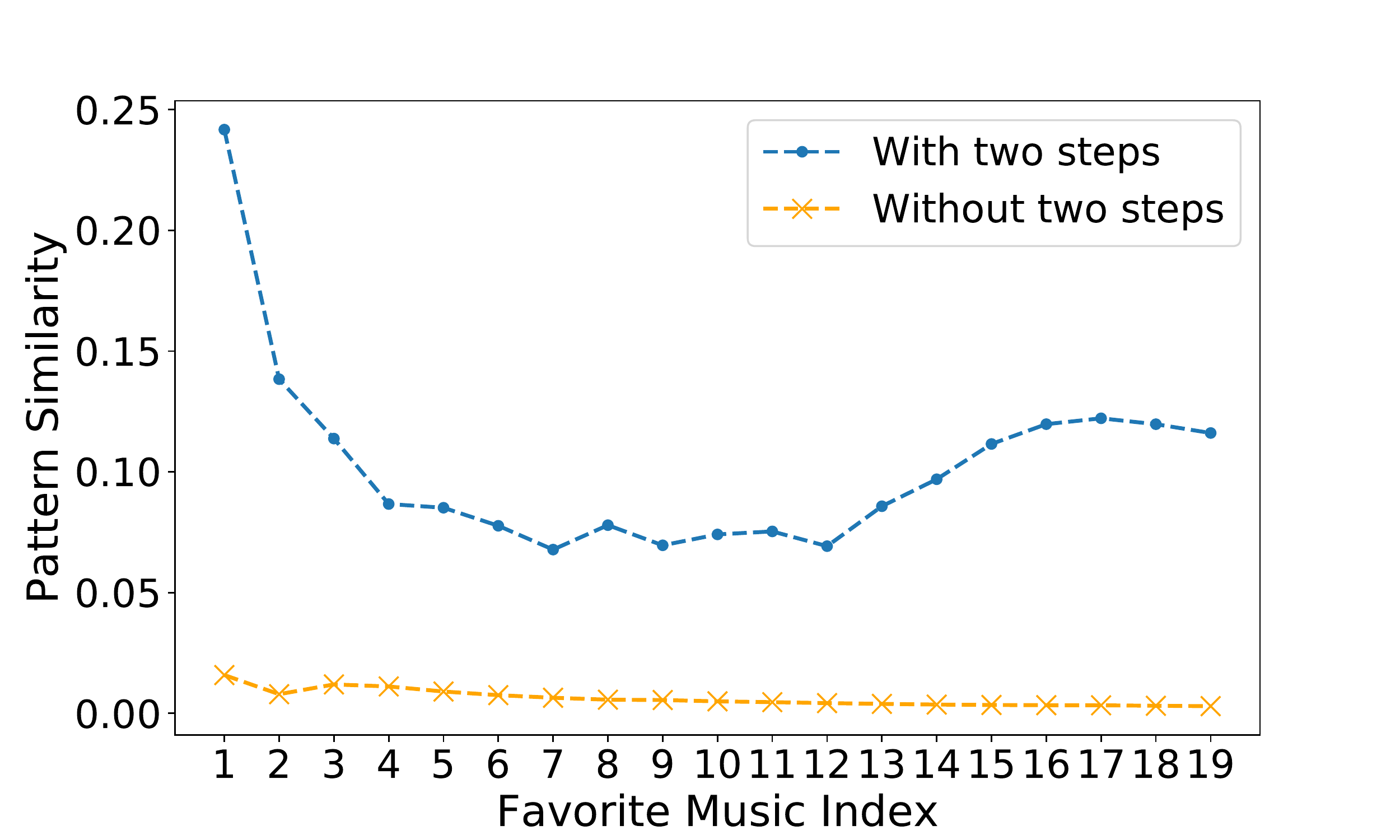}}	
		\subfigure[p = 5]
		{\includegraphics[width=1.7in,trim=10 0 40 50,clip]{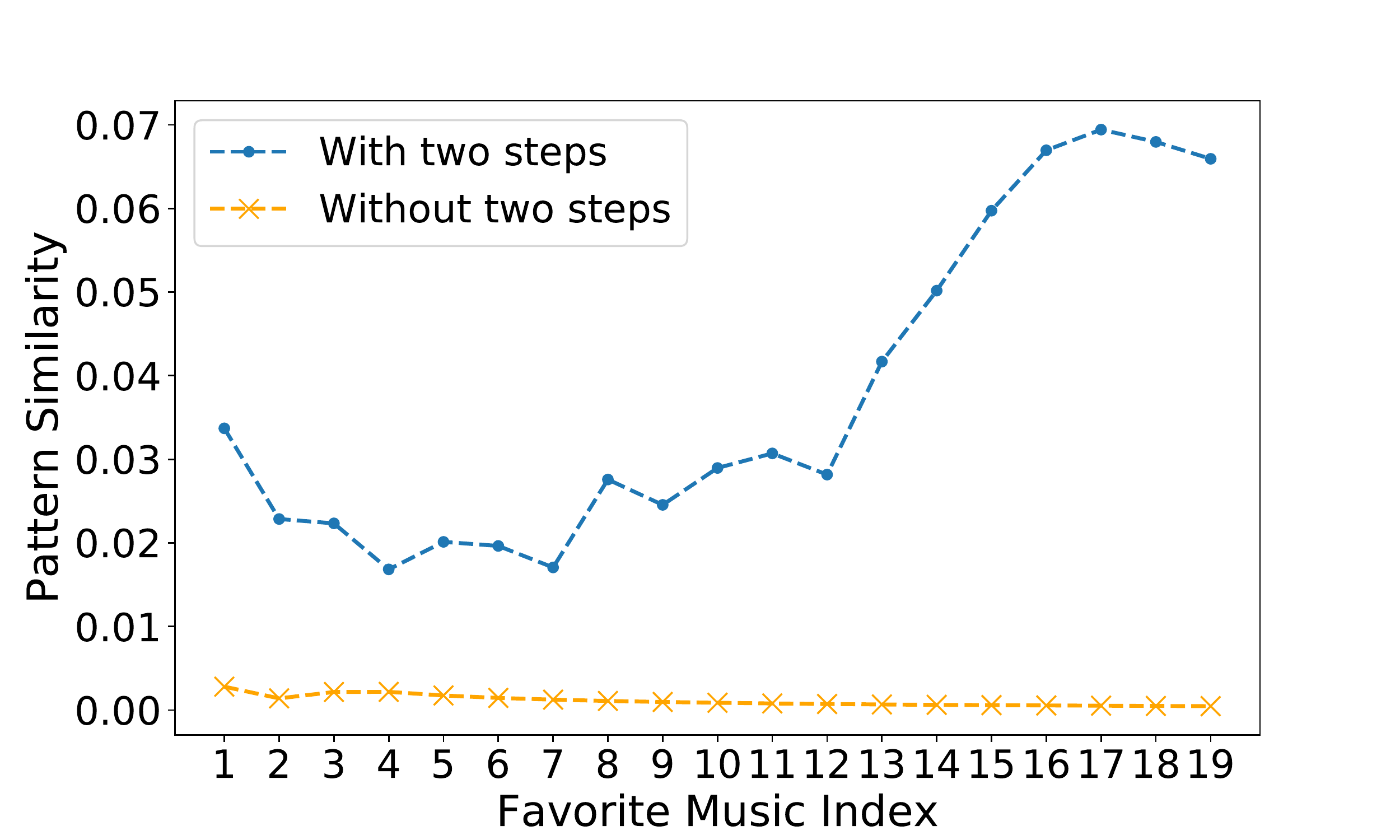}}
		\caption{Pattern similarity for 19 pieces of the user’s favorite music when different music pattern lengths $p$ are chosen.}
		\label{fig4}
		
	\end{figure*}
	To verify that the proposed two steps in the transfer phase enhance output performance, we calculated pattern similarity before and after implementing both steps. Fig. \ref{fig4} depicts the pattern similarity of the transferred music and 19 pieces of a user’s favorite music before and after implementing the two steps. Music pattern lengths $p$ from 2 to 5 were calculated. First, a smaller music pattern length shows higher pattern similarity. This result is understandable because a match between transferred music and a user’s favorite music is more likely if the pattern length is short. Second, implementing two steps improved pattern similarity in all music samples, indicating that these steps increase the pattern similarity of transferred music and the user’s favorite music.

	\begin{figure*}
		\subcapcentertrue
		\subfigure[input music A]
		{\includegraphics[width=1.76in,height=1.7in,trim=10 40 20 40]{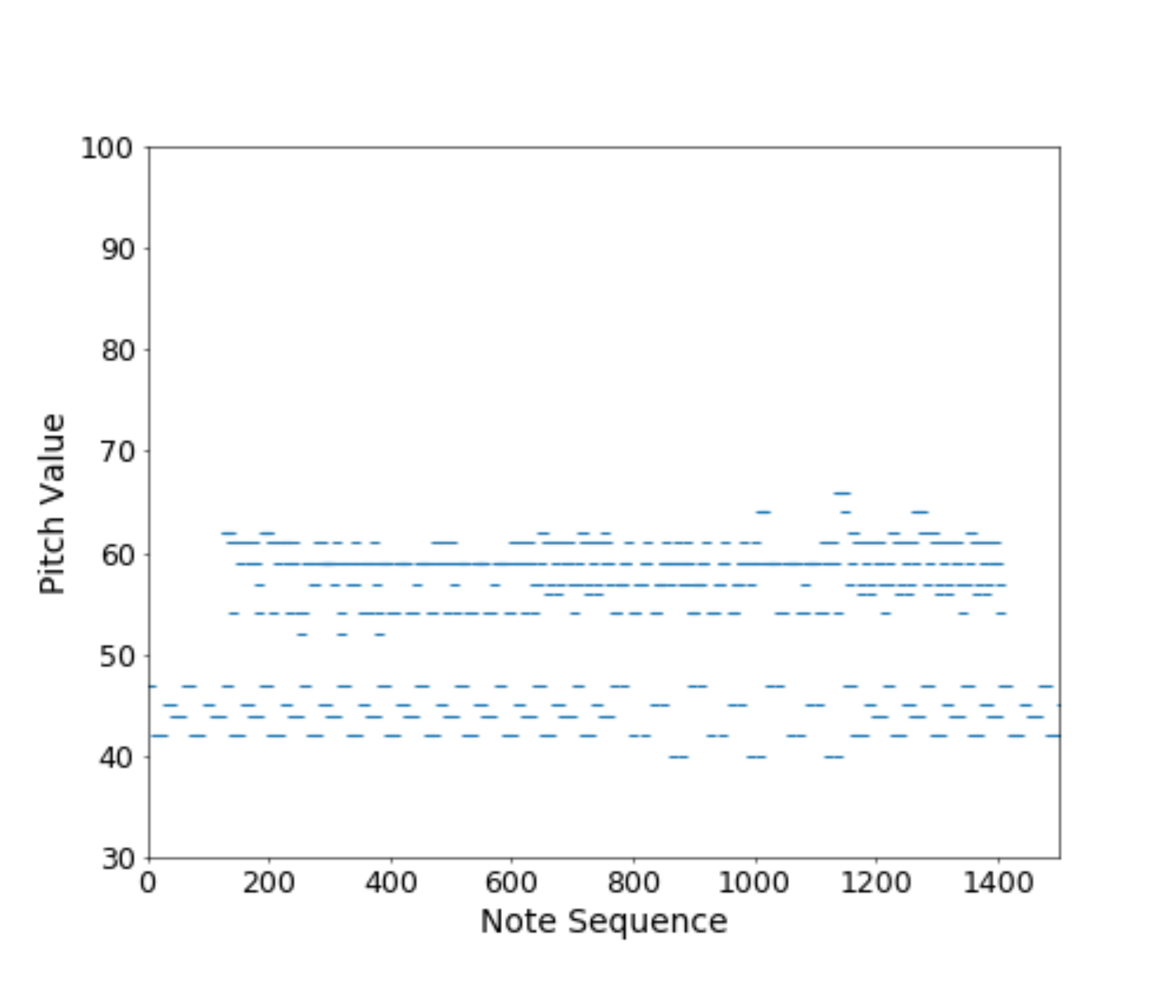}}
		\subfigure[input music B]
		{\includegraphics[width=1.76in,height=1.7in,trim=10 40 20 40]{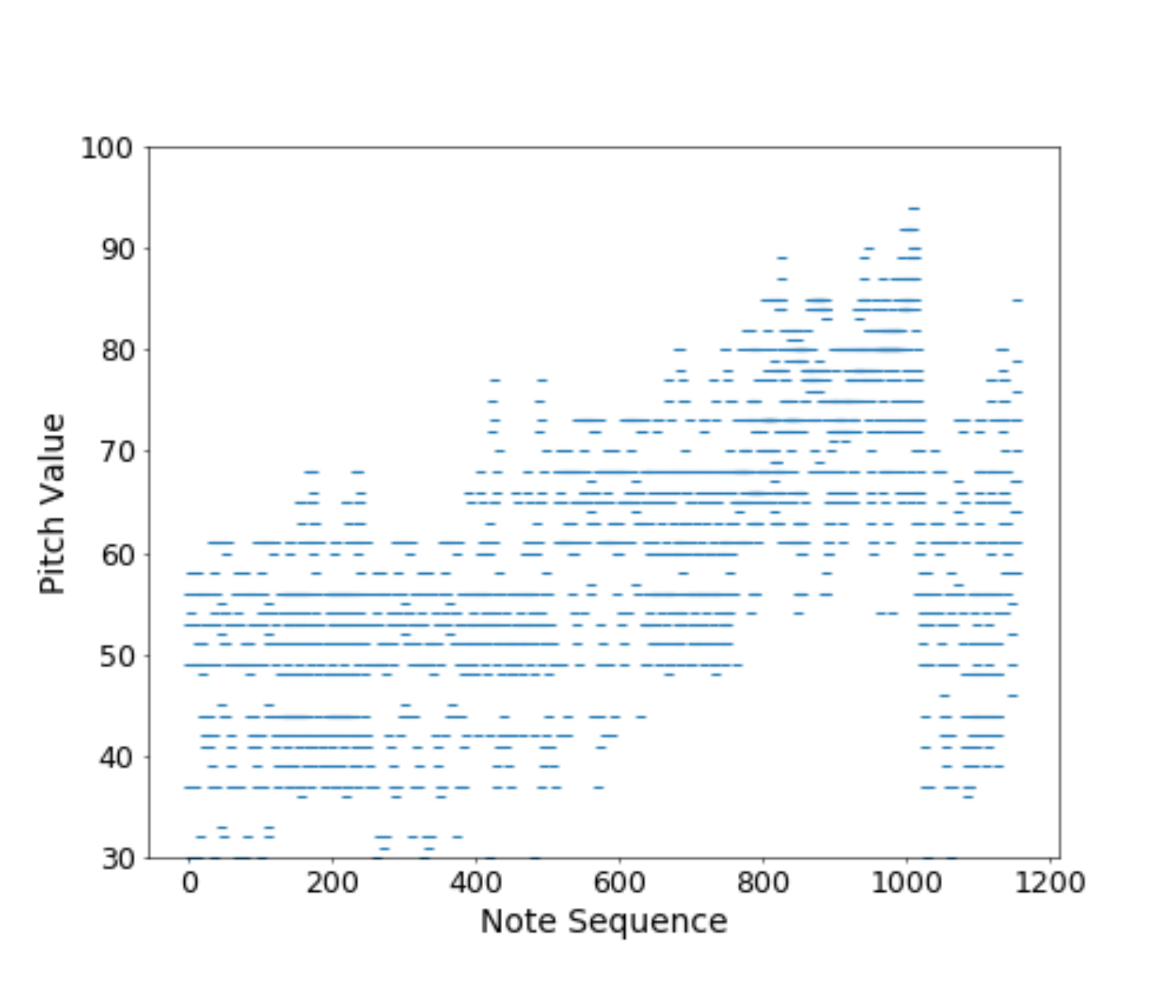}}
		\subfigure[user’s favorite music A]
		{\includegraphics[width=1.76in,height=1.7in,trim=10 40 20 40]{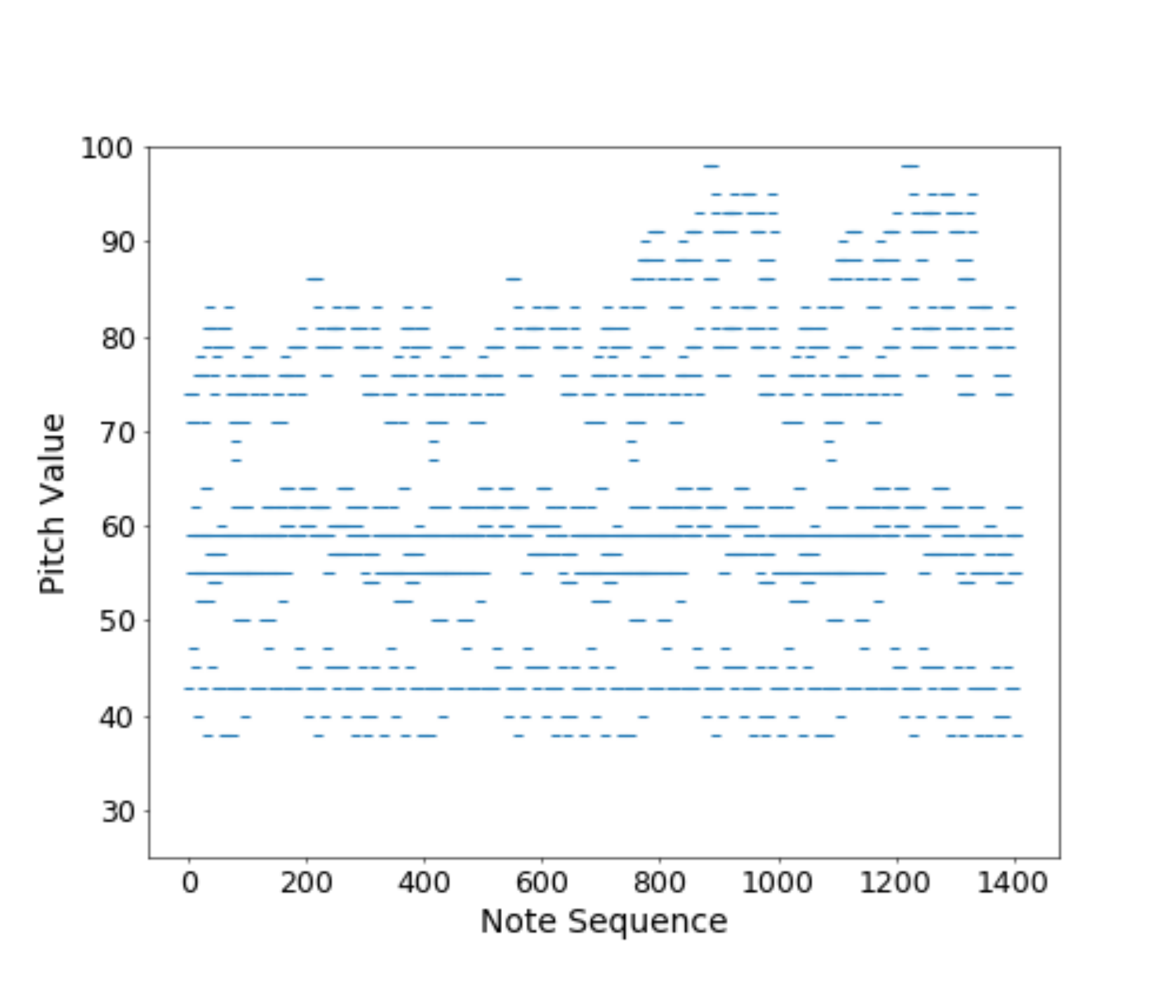}}
		\subfigure[user’s favorite music B]
		{\includegraphics[width=1.76in,height=1.7in,trim=10 40 20 40]{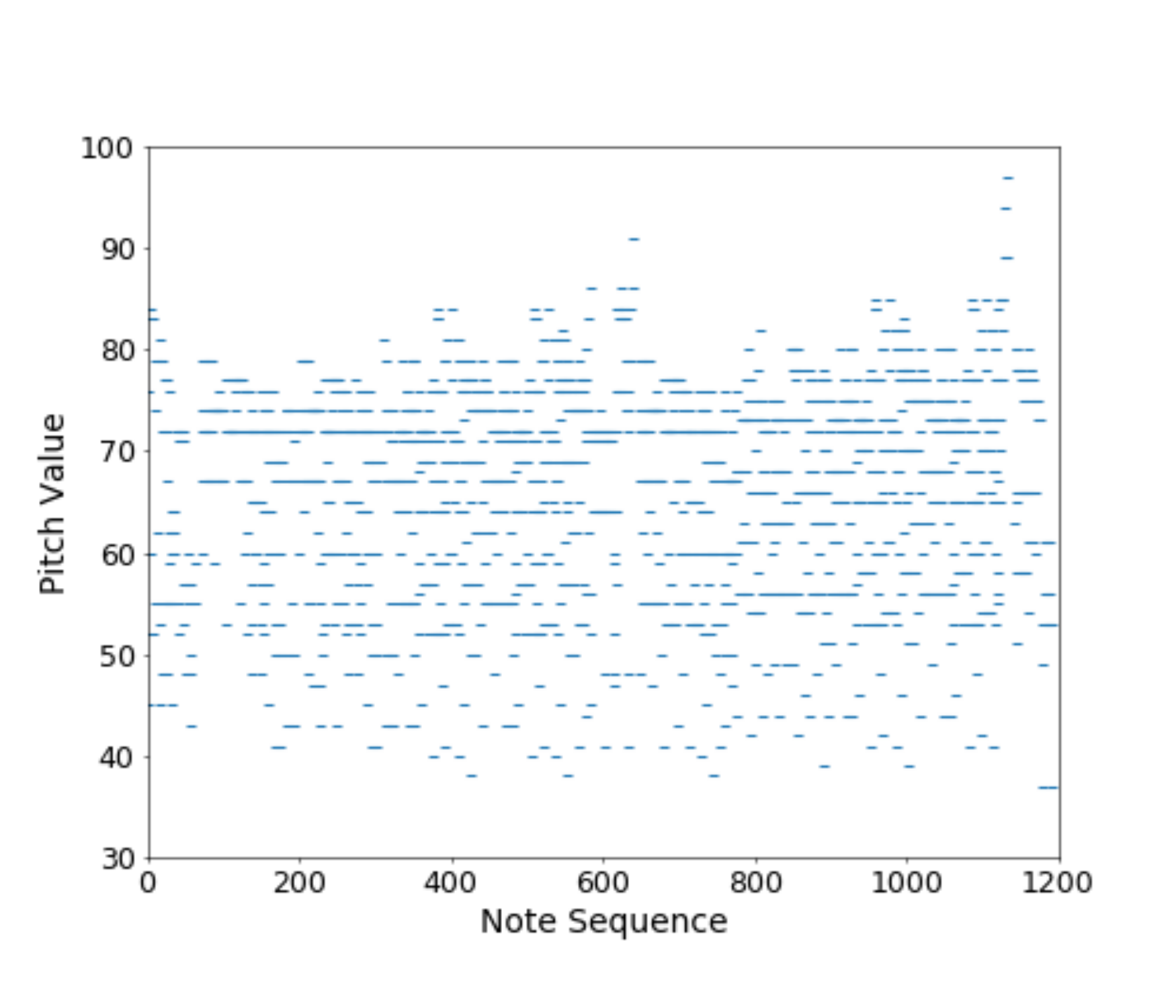}}
		\subfigure[transferred music A: input music A + user’s favorite music A]
		{\includegraphics[width=1.76in,height=1.7in,trim=10 40 20 20]{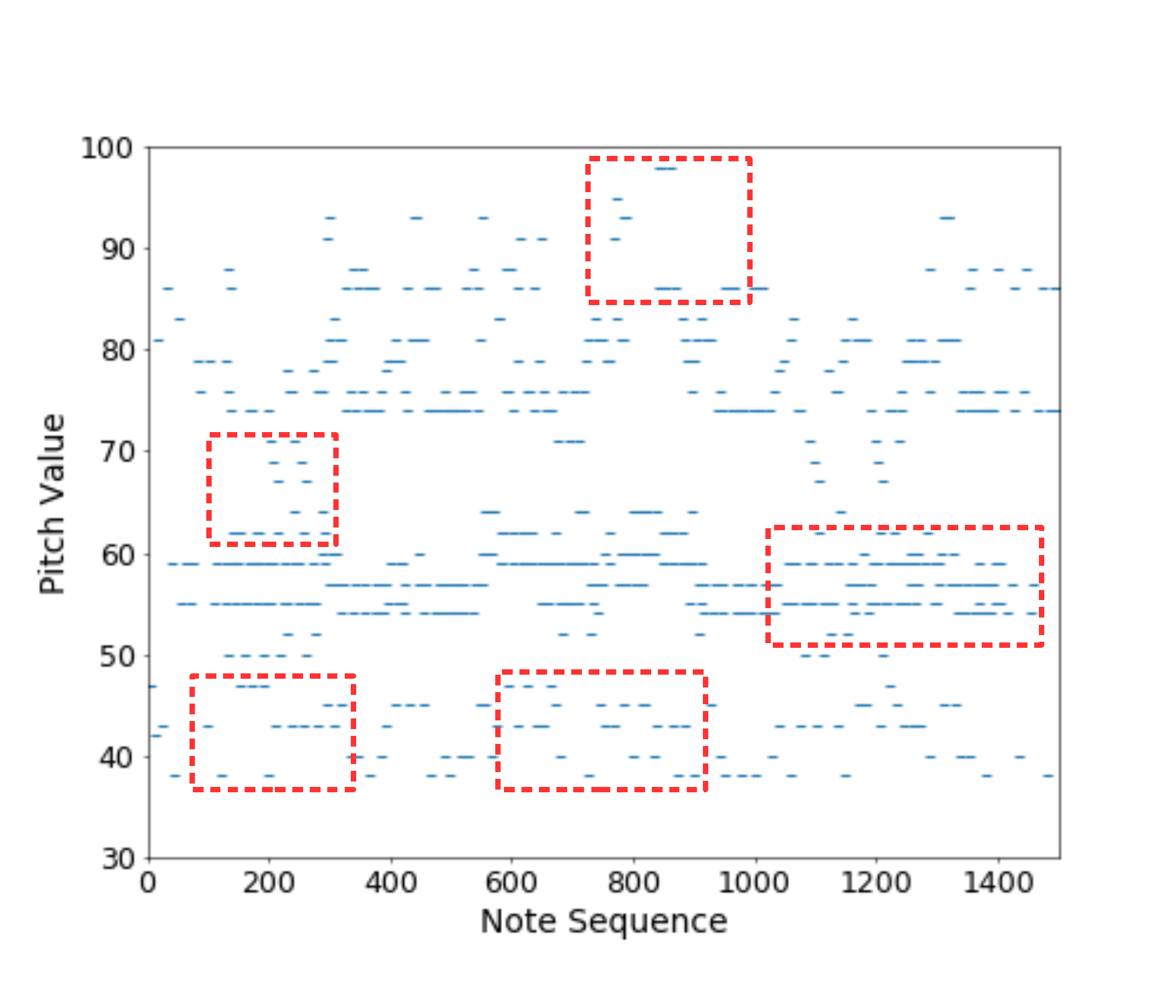}}
		\subfigure[transferred music B: input music A + user’s favorite music B]
		{\includegraphics[width=1.76in,height=1.7in,trim=10 40 20 20]{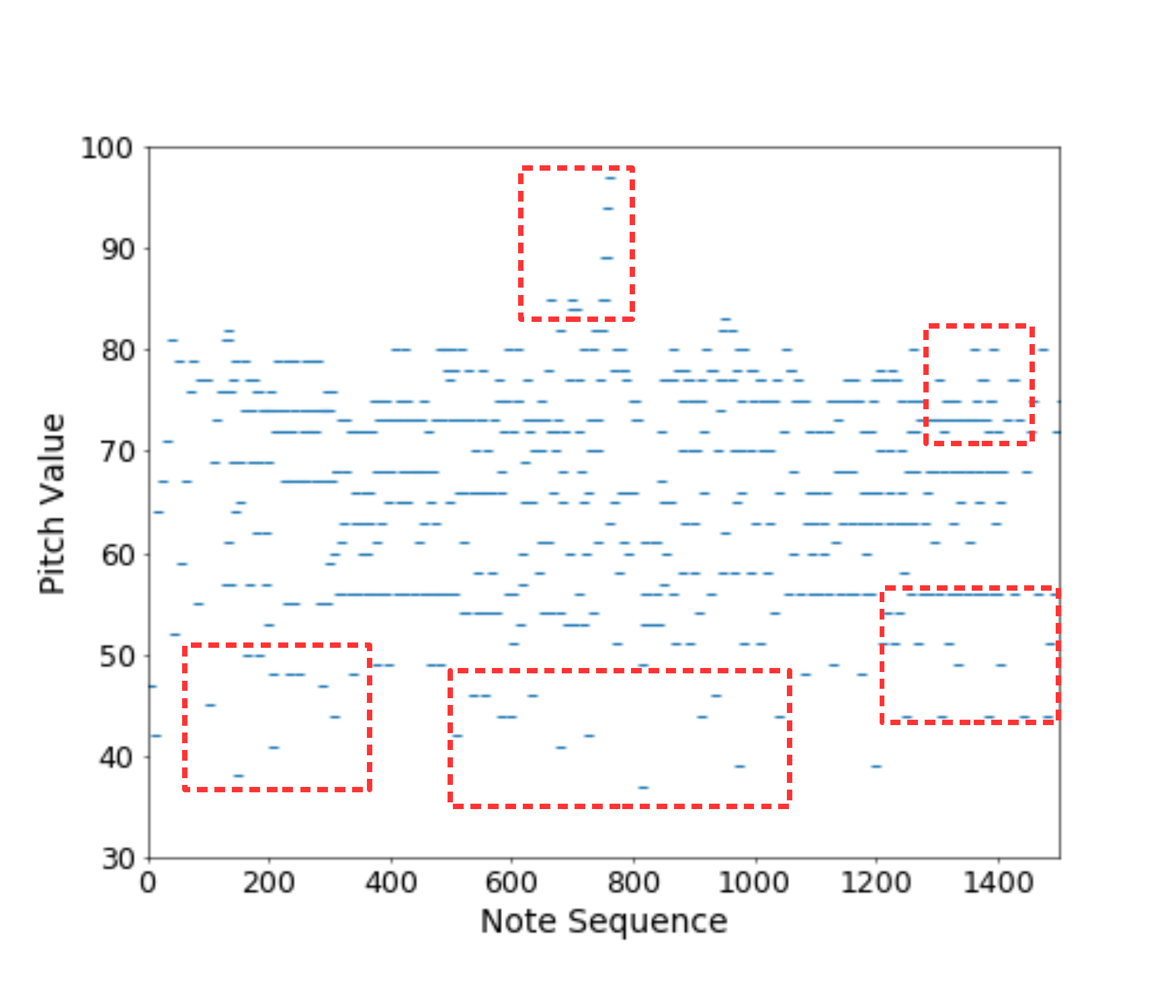}}
		\subfigure[transferred music C: input music B + user’s favorite music A]
		{\includegraphics[width=1.76in,height=1.7in,trim=10 40 20 20]{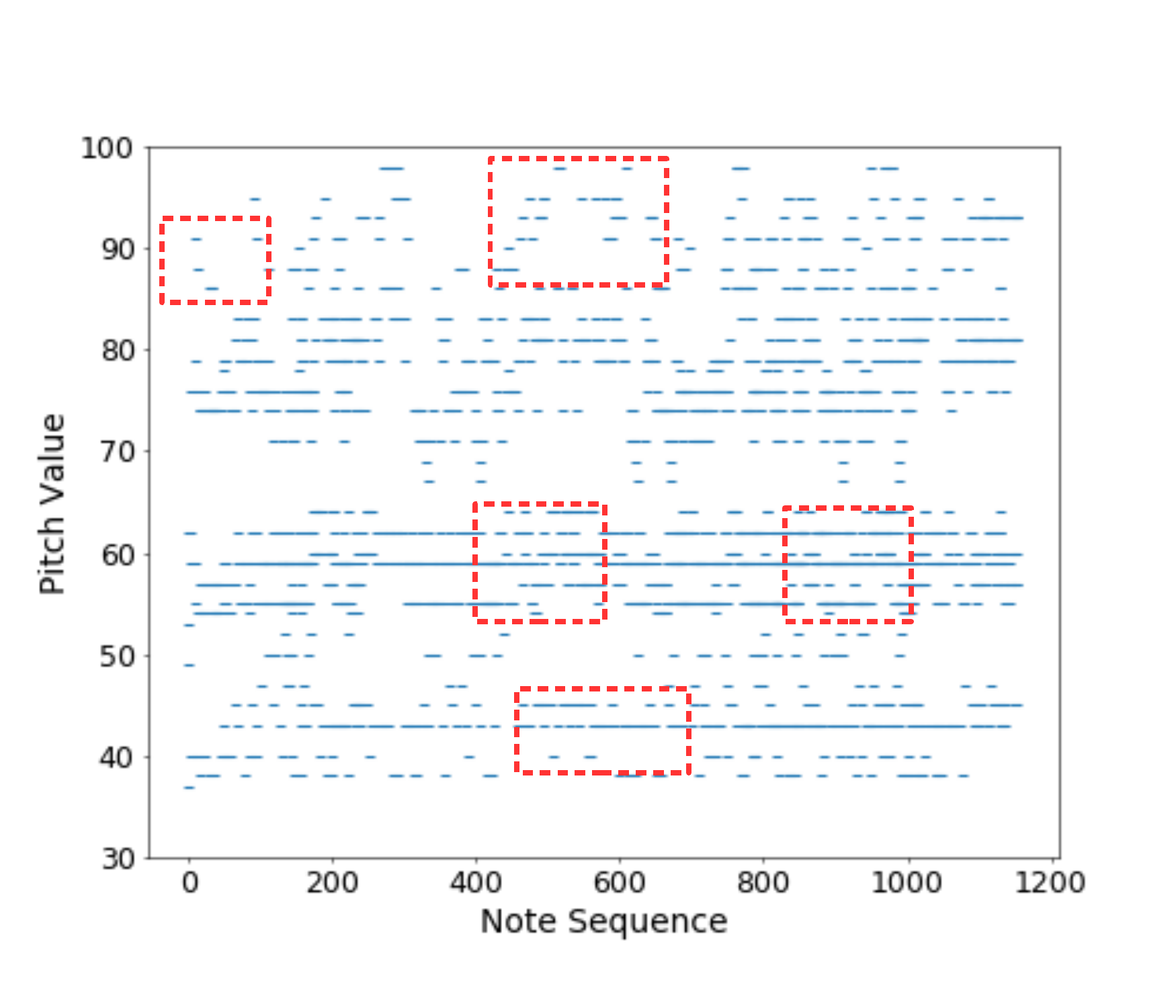}}
		\subfigure[transferred music D: input music B + user’s favorite music B]
		{\includegraphics[width=1.76in,height=1.7in,trim=10 40 20 20]{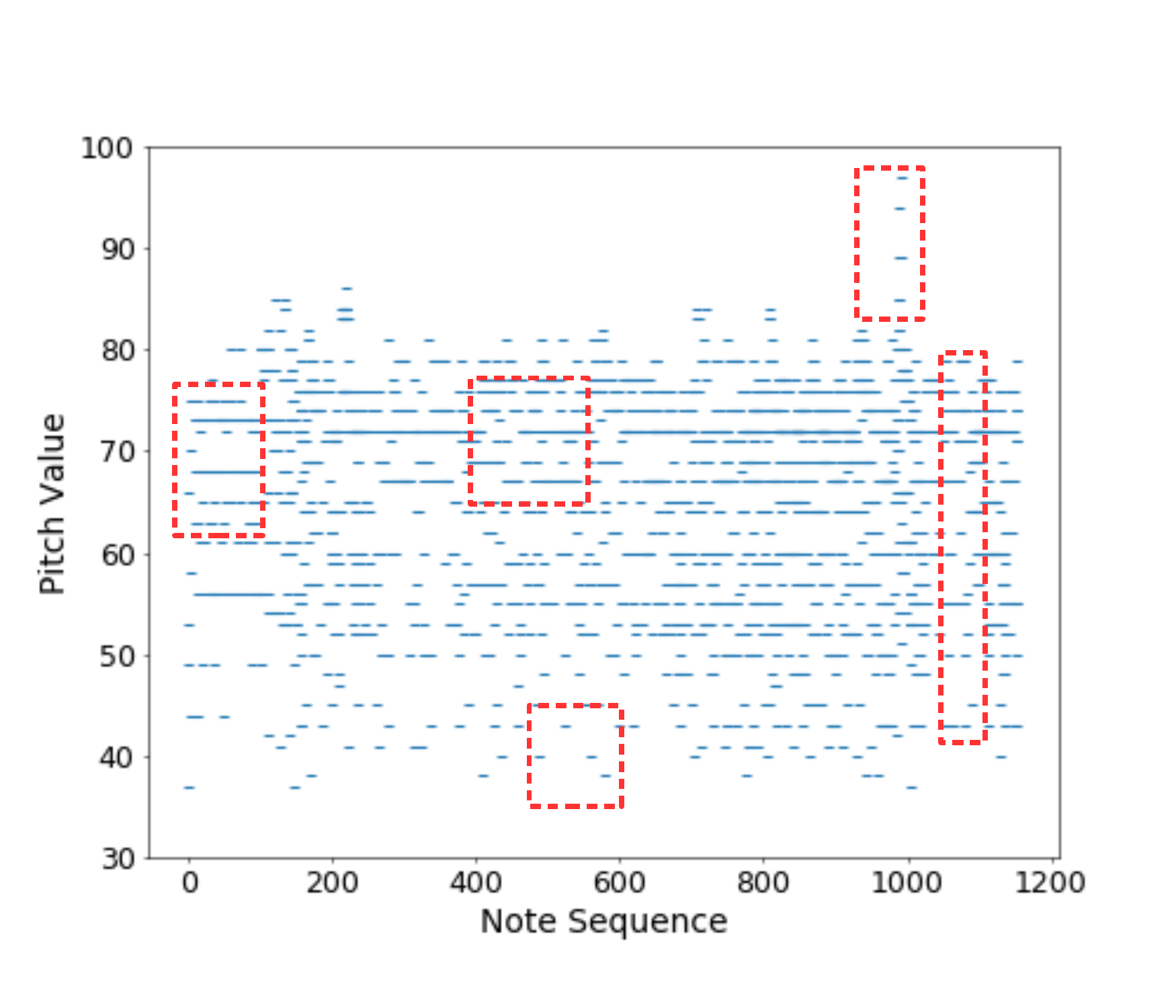}}		
		\caption{Demonstration of music pattern similarity when music pattern length $p=4$. Red rectangle indicates that the pattern matches the melody in the user’s favorite music.}\label{fig:1}
	\end{figure*}

	In addition, in Table \ref{tab21}, $D_{P}$ and $D_{N}$ of UP-Transformer increase significantly compared to the other two models and input; therefore, implementing two music steps significantly enhances the similarity in pitch value between the new music and the user’s favorite music. However, $D_{D}$ and $D_{IOI}$ do not change much; only \textit{Note On} events were transferred in the experiment. This result satisfies our goal of acquiring knowledge from the user’s favorite music (\textit{Note On}) while retaining other information from the input music (e.g., \textit{Note Duration}, \textit{Position}, and so on.). 
	
	Fig. \ref{fig:1} displays two randomly selected input examples, two examples of the user’s favorite music, and four transferred samples with music pattern similarity when the music pattern length $p=4$. Note duration and instrument information are each omitted from this figure for simplicity. In the transferred music, some music patterns that match the user’s favorite music are enclosed in a red rectangle. First, this figure indicates that the transferred music maintains a similar structure as the input music given that the note position, note numbers, and total length of the note sequence remain the same. Second, the melody of the transferred music tends to contain the melody of a user’s favorite music. For example, notes center around 80, 60, and 40 in the user’s favorite music A. The pitch values of transferred music A and C also center around 80, 60, and 40. The pitch values of transferred music B and D are scattered but are similar to the pitch values of the user’s favorite music B. Third, red rectangles appear throughout the music piece, showing that the entire piece contains the melody of a user’s favorite music instead of mere clips.

	\subsection{Correlation test}
	To validate that the proposed metric, PS, can measure melodic similarity and that the result is consistent with human judgment, we referred to the Kendall correlation to quantify the relationship between PS and the qualitative similarity score. This correlation can be used on small data samples and does not make assumptions about the underlying distributions, noise, or nature of relationships \cite{dobrian2011understanding}. If the music has a high PS value, then the music will have a high qualitative similarity score as PS reflects melodic similarity, and the Kendall correlation will be high. TABLE \ref{tab10} shows the Kendall correlation and the p-value when a different music pattern length $p$ is chosen. Results reflect a positive correlation between PS and human judgment in terms of music similarity when $p=2,3,4$. This pattern also indicates that humans can recognize the similarity in two pieces of music even if the music pattern length is quite short. Although the p-value is not significant when $p=5$, it shows a positive relationship (Kendall correlation $=0.22$) between PS and the qualitative similarity score.  
	
	\begin{table}
		\setlength\tabcolsep{12pt}
		\centering
		\caption{Kendall correlation between pattern similarity and the qualitative similarity score when a different music pattern length $p$ is chosen.}
		\begin{tabular}{ccc}
			\toprule
			Music Pattern Length &Kendall correlation&P-value \\
			\midrule
			$p=2$ &0.64&0.0005\\
			$p=3$  &0.80&0.0000\\
			$p=4$  &0.72&0.0001\\		
			$p=5$&0.22&0.2199\\	 
			\bottomrule
		\end{tabular}
		\label{tab10}
	\end{table}

	\section{Conclusion}
	We constructed a new UPMT problem and proposed a hybrid model that combines Transformer-XL with music knowledge based on only one piece of a user’s favorite music. The novel favorite-aware loss function and two steps allows this model to personalize input music for different users. We also proposed a new metric to measure PS between two pieces of music. Quantitative and qualitative experiments confirmed that UP-Transformer can transfer input music to fit users’ preferences with high quality. 
	
	We will focus on several research directions in the future. First, to increase the diversity of music and explore more music transfer possibilities to fit users’ preferences, we plan to transfer multiple music events in sequence at the same time and to consider different music features such as lyrics and performance. Second, to improve model performance, different music theories can be applied across two steps (e.g., considering both melodic intervals and harmonic intervals). Third, to enhance model efficiency, a one-size-fits-all model that suits all users’ preferences can be introduced to decrease the fine-tuning time. Codes and listening examples are available online at https://github.com/zhejinghu/UPMT.

	\appendices

	\section*{Acknowledgment}

	This study was funded by P0030934 DaSAIL-Music Generation via Machine Composition.	
	
	\ifCLASSOPTIONcaptionsoff
	\newpage
	\fi

	
	
	\bibliographystyle{IEEEtran}
	\bibliography{IEEEabrv,bare_jrnl}
	
	%
	
	%

	\begin{IEEEbiographynophoto}{Zhejing Hu }
		is currently a Ph.D. candidate with the Department of Computing in The Hong Kong Polytechnic University. His research interests include music style transfer, algorithmic composition, artificial intelligence, and machine learning.

	\end{IEEEbiographynophoto}
	
	\begin{IEEEbiographynophoto}{Yan Liu }
		obtained Ph.D. degree in Computer Science from Columbia University in the US. In 2005, she joined The Hong Kong Polytechnic University, Hong Kong, where she is currently an Associate Professor with the Department of Computing and the director of Cognitive Computing Lab. Her research interests span a wide range of topics, ranging from brain modeling and cognitive computing, image/video retrieval, computer music to machine learning and pattern recognition.
	\end{IEEEbiographynophoto}
	
	\begin{IEEEbiographynophoto}{Gong Chen }
		received the Ph.D. degree in artificial intelligence from The Hong Kong Polytechnic University. He is currently a Post-Doctoral Researcher at Shenzhen Research Institute, The Hong Kong Polytechnic University. His research interests include artificial intelligence, computer music, brain imaging, and affective computing.
	\end{IEEEbiographynophoto}

	\begin{IEEEbiographynophoto}{Yongxu Liu }
		is currently a Ph.D. candidate with the Department of Computing in The Hong Kong Polytechnic University. Her research interests include noisy and imbalance data, data preprocessing, data mining, machine learning, and deep learning.
	\end{IEEEbiographynophoto}

\end{document}